%% file: main.tex
\documentclass[11pt]{article}

\usepackage[final]{acl}

\usepackage{times}
\usepackage{latexsym}
\usepackage{hyperref}  %
\usepackage[T1]{fontenc}
\usepackage[utf8]{inputenc}
\usepackage{microtype}

\usepackage{inconsolata}
\usepackage{amsmath, amssymb, amsfonts}
\usepackage{graphicx}
\usepackage{tcolorbox}
\usepackage{graphicx}
\usepackage{algorithm}
\usepackage{algorithmic} 
\usepackage{booktabs}
\usepackage{multirow}
\usepackage{makecell}
\usepackage{xcolor}
\usepackage{lipsum} % 仅用于示例文本，实际使用时可以删除
\usepackage{arydshln}
\tcbuselibrary{listings, breakable}
\usepackage{enumitem}
\definecolor{lavenderback}{RGB}{252, 252, 253} % 近白背景
\definecolor{lavenderframe}{RGB}{102, 78, 145} % 紫边框
\definecolor{codeback}{RGB}{245, 246, 248}     % 浅灰代码块

\usepackage{varwidth}
\usepackage{enumitem}
\usepackage{caption}
\usepackage[table]{xcolor}
\usepackage{pifont}
\newcommand{\xmark}{\ding{55}}
\title{UniSRM: A Unified Speech Reward Model for Reasoning-Based Fine-grained Assessment}

\author{
 \textbf{Yuanyuan Wang\textsuperscript{1}},
 \textbf{Dongchao Yang\textsuperscript{1}},
 \textbf{Yayue Deng\textsuperscript{1}},\\
 \textbf{Zhiyong Wu\textsuperscript{1,2,$\dagger$}},
 \textbf{Yiwen Guo\textsuperscript{3}},
 \textbf{Helen Meng\textsuperscript{1}},
 \textbf{Xixin Wu\textsuperscript{1,$\dagger$}},
\\
\\
 \textsuperscript{1} The Chinese University of Hong Kong,
 \textsuperscript{2} Tsinghua University,
 \textsuperscript{3} Independent Researcher
}

\begin{document}
\maketitle
{\renewcommand{\thefootnote}{}\footnotetext{$^\dagger$ Corresponding authors.}}
\begin{abstract}
% Existing speech generation optimization is bottlenecked by reward models that are opaque in decision-making rationale, single-dimensional, and human-intensive. 
% Motivated by the observation that speech understanding is often easier to elicit than high-quality generation, 
% In this study, we explore whether large speech-language models can serve as effective reward models via comprehension and reasoning. 

% Existing speech reward modeling is hindered by opaque scoring, incomplete evaluation dimensions, and insufficient supervision over the reasoning process, which together limit reliable optimization for speech generation. 
% To address these issues, we propose UniSRM, a unified speech reward model that produces multi-dimensional, interpretable reward signals with reliable reasoning. 
% We first construct UniSRM-Data, a dataset covering four evaluation tasks ranging from utterance-level quality to context-level coherency. 
% We also construct UniSRM-Bench, a comprehensive benchmark across diverse tasks. 
% We then train UniSRM to provide dimension-wise evidence for its judgments. 
% To improve reasoning reliability, we further introduce a fine-grained optimization strategy that stabilizes the reasoning process and mitigates inconsistencies between rationales and final decisions. 
% Experiments show that UniSRM yields reliable judgments and serves as an effective training signal for improving speech quality.

Evaluating speech generation still relies heavily on human judgments, such as Mean Opinion Score (MOS), which are expensive, subjective, and difficult to reproduce at scale. 
While a few recent studies have begun to explore AudioLLM-based judge models, existing efforts typically target only a narrow set of scenarios (e.g., utterance-level quality or single-turn dialogue) and provide limited coverage of diverse speech generation tasks and evaluation dimensions. 
In this work, we propose UniSRM, a unified speech reward model that can support multi-dimensional, interpretable reward signals with reliable reasoning.
To support training and evaluation, we introduce UniSRM-Data and UniSRM-Bench, covering speech evaluation tasks from utterance-level quality to context-level coherence. 
Based on this dataset, we present the unified speech reward model, UniSRM, with a two-stage pipeline that enables reasoning-based fine-grained assessment.
Furthermore, we introduce Reasoning-Consistent Rewards to improve the reliability of the reasoning process.
Experiments show that UniSRM delivers more reliable and human-aligned judgments across a broad range of speech evaluation tasks, offering a practical foundation for scalable and unified evaluation of speech quality~\footnote{
The checkpoint and dataset are publicly available at \href{https://github.com/lavendery/UniSRM}{https://github.com/lavendery/UniSRM}.}. 

\end{abstract}

\input{content/01intro}
\input{content/02related}
\input{content/03dataset}

\input{content/03method}

\input{content/04exp}

\input{content/05_conclusion}

\bibliography{custom}

\newpage
\appendix
\input{appendix/system_prompt}

\input{appendix/evaluation_dimension}
\input{appendix/response_demo}

\input{appendix/model_config}
\input{appendix/dataset}

\input{appendix/eg_exp}
\input{appendix/human_annotation}
\input{appendix/multi-turn}
\input{appendix/input-parity}

\end{document}

%% file: content/01intro.tex
\section{Introduction}

Autoregressive large language models (LLMs) have revolutionized natural language processing with their strong generation and reasoning capabilities~\cite{brown2020gpt3, touvron2023llama, team2023gemini}.
Reinforcement learning (RL) \citep{ouyang2022training, schulman2017proximal} has emerged as a promising framework for improving model alignment with human preferences through reward-based feedback. 
However, in speech generation, the lack of well-designed reward models makes optimization particularly challenging. 
The Mean Opinion Score (MOS) \citep{ITU-T_P800_1996} is an ideal, widely accepted criterion for assessing speech quality. However, it is costly to collect and inherently subjective, with ratings varying across listeners and datasets, which lack reliable optimization targets and hinder reproducible comparisons.

Existing speech generation methods powered by RL mainly rely on two categories of reward signals. 
First, some works~\citep{zhang2024speechalign, fu2024asrrl, hu2024robust, sun2025f5r, chen2025fine} utilize classical speech reward signals, which are typically objective metrics such as WER, speaker similarity (SIM), and UTMOS\citep{saeki2022utmos}. 
While these metrics provide effective indicators, each of them captures only a \textit{single aspect} of speech.
For instance, WER mainly reflects textual correctness, and SIM measures timbre similarity to a reference speaker.
As a result, such rewards cannot holistically evaluate speech.
Moreover, when used as reward signals, they are often treated as black-box scorers that output a single scalar value without intermediate and explicit explanations. 
This \textit{lack of transparency} may introduce bias and inconsistency into the optimization process, ultimately undermining the reliability of reward-based alignment for speech generation.
Secondly, some works use Large Audio Language Models (LALMs) as speech judges. 
For example, WavReward~\cite{ji2025wavreward} and SageLM~\cite{ge2025sagelm} fine-tune LALMs for single-turn spoken dialogue evaluation.
SpeechJudge~\cite{zhang2025speechjudge} trains a generative reward model largely centered on naturalness over utterance-level speech.
These works remain \textit{limited in task coverage}, typically focusing on only utterance-level speech or single-turn dialogue. 
They also have incomplete evaluation dimensions, such as overlooking speaker similarity.
Moreover, the rule-based RL~\cite{ge2025sagelm} provides limited supervision over reasoning, which cause inconsistency between the generated rationales and final decisions.
Overall, current research on speech reward modeling still faces four main challenges: (1) lack of transparency in scoring; (2) incomplete evaluation dimensions; (3) limited task coverage; and (4) insufficient supervision over the reasoning process.

To address these limitations, we propose UniSRM, a unified speech reward model designed to produce multi-dimensional, interpretable reward signals backed by reliable reasoning.
Experiments demonstrate that UniSRM yields judgments that are not only more reliable but also better aligned with human preferences.
Our main contributions are summarized as follows:
\begin{itemize}[topsep=0pt, itemsep=0pt, parsep=0pt, partopsep=0pt]
    \item \textbf{Comprehensive Data and Benchmark.} We construct \textsc{UniSRM-Data}, a unified dataset that covers speech evaluation tasks \textit{from utterance-level quality to context-level coherency}.
    In parallel, we introduce \textsc{UniSRM-Bench} as a comprehensive benchmark for unified speech reward modeling.

    \item \textbf{A Unified Reward Model with Explicit Decomposition.} Building on \textsc{UniSRM-Data}, we develop \textsc{UniSRM} with a two-stage training pipeline, which explicitly decomposes speech assessment into multiple dimensions, enabling fine-grained evaluation across diverse tasks.
    
    \item \textbf{Reasoning-Consistent RL Optimization.} During the RL stage, we propose RCR-GRPO (Reasoning-Consistent Rewards), which assigns rewards at the dimension-wise reasoning process to improve reliability. 

\end{itemize}

%% file: content/02related.tex
\section{Related Work}
\begin{figure*}[t]
    \centering
    % \vspace{-5mm}
    \includegraphics[width=1.0\textwidth]{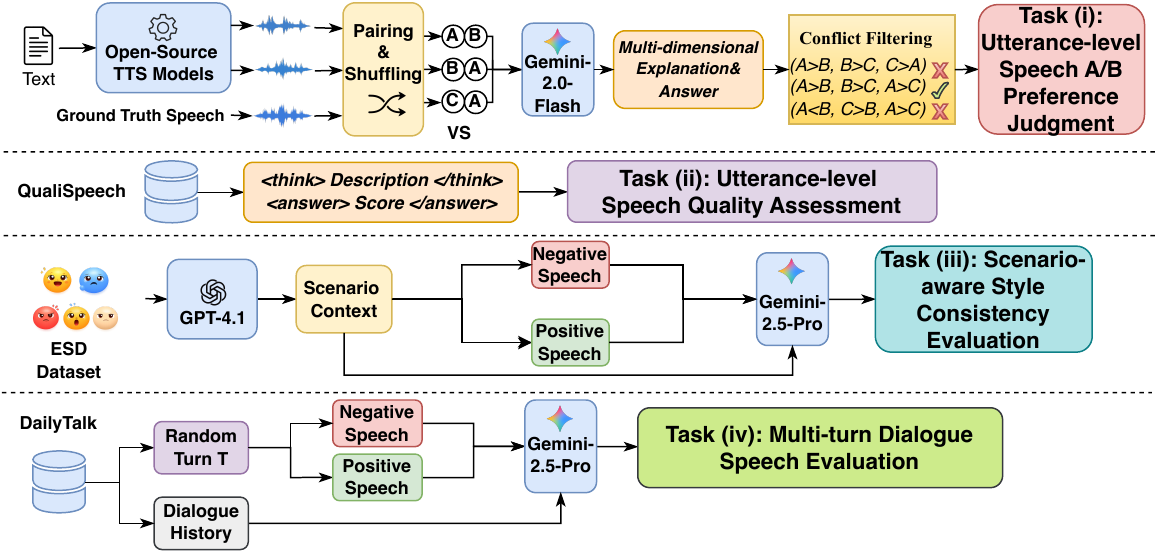}
    \vspace{-20pt}
    \caption{The Pipeline of UniSRM Dataset construction.} 
    \label{fig:data_outline}
    \vspace{-15pt}
\end{figure*}

\subsection{Multimodal Reward Models}

Reinforcement learning from human feedback (RLHF) and its variants have become an effective paradigm for aligning large language models (LLMs) with human preferences~\citep{ziegler2019fine,ouyang2022training,rafailov2023direct}. Recent works extend this paradigm to multimodal settings, where reward models provide supervisory signals for both understanding and generation tasks over image, video, and audio~\citep{team2023gemini,lee2023aligning, yang2023diffsound,wang2024lift,wang2024consistent, yang2024uniaudio,yang2025simplespeech,liu2025improving,zhao2025omnialign,wang2025unified, wang2025unisep,wang2025audiocomposer,yang2025humanomniv2,yang2026uniaudio}.
However, despite this progress in multimodal settings, reward modeling for speech still has meaningful room for further improvement~\citep{yang2024uniaudio15, yang2025almtokenizer}.

\subsection{Speech Reward Models}
Using Large Audio Language Models (LALMs) as automated speech judges has recently received growing attention.
ATT~\cite{wang2025audio} and ALLD~\cite{chen2025audio} both introduce a human-likeness speech evaluation corpus and train LALMs to describe and score speech quality in a human-aligned manner. 
QualiSpeech~\cite{wang2025qualispeech} develops a detailed dataset for low-level speech quality assessment. 
AudioJudge~\cite{manakul2025audiojudge} explores prompting strategies to elicit multi-aspect judgments.
WavReward~\cite{ji2025wavreward} extends LALMs to evaluate both IQ and EQ for spoken dialogue systems, but is restricted to single-turn dialogue. SageLM~\cite{ge2025sagelm} also trains an end-to-end spoken dialogue evaluator via SFT for single-turn conversational quality. 
SpeechLLM-as-Judges~\cite{wang2025speechllm} fine-tunes a speech quality LLM on a large-scale SpeechEval dataset to perform assessment, comparison, improvement suggestion, and deepfake detection. 
SpeechJudge~\cite{zhang2025speechjudge} trains a generative reward model, aiming at utterance-level preference evaluation over paired speech samples.

Some of these above approaches offer evaluation dimensions that are not sufficiently fine-grained and comprehensive~\cite{wang2025audio,chen2025audio,wang2025qualispeech}, while others inherently inherit the limited understanding capacity of the underlying LAMs and may therefore produce shallow judgments~\cite{manakul2025audiojudge}, and some are restricted by insufficient coverage of task scenarios~\cite{wang2025speechllm,wang2026dualspeechlm,zhang2025speechjudge,ge2025sagelm,ji2025wavreward}. 
Moreover, rule-based reinforcement learning lacks supervision over the reasoning process, which can lead to inconsistency between the rationale and final result~\cite{ge2025sagelm, zhang2025speechjudge}. 
% Overall, existing audioLLM-based judges still have substantial room for improvement.

In this paper, we propose an end-to-end speech reward model, UniSRM, that decomposes speech quality into multiple complementary dimensions and generates explicit reasoning traces before producing an aggregated preference decision, thus offering both a richer supervision signal and improving interpretability.

%% file: content/03dataset.tex
\section{UniSRM-Data and UniSRM-Bench}
\label{sec:data}
As shown in Figure~\ref{fig:data_outline}, we construct a unified dataset that covers speech evaluation tasks ranging
\textit{from utterance-level quality to context-level coherency}, including:
(i) utterance-level speech A/B preference judgment,
(ii) utterance-level speech quality assessment,
(iii) scenario-aware style coherency evaluation conditioned on textual context, and
(iv) multi-turn dialogue speech evaluation conditioned on dialogue history. 
In the following, we describe the construction process of each task in detail.

\paragraph{Task 1: Utterance-Level Speech A/B Preference Judgment.}
Pairwise preference comparison is often easier and more reliable than absolute scoring for speech quality assessment, as it directly asks which of two speech samples of the same target text is better overall. Such relative ranking, compared to absolute scores, is less susceptible to variations between judges.

Given a text prompt $x$, we use multiple open-source TTS models $s$ to synthesize diverse candidate speech signals.
\begin{equation}
\label{eq1}
    \mathcal{S}(x)=\{ s_{1}, s_{2}, \dots, s_{K} \},
\end{equation}
where $K$ is the number of open-source TTS models.
These speech samples, generated from the same textual content by different synthesis models, are paired to form comparison candidates. 
We additionally include the ground-truth recording $S^{\text{GT}}$ as another candidate. We then form unordered comparison pairs:
\begin{equation}
\begin{aligned}
\mathcal{P}(x)
= \Bigl\{ (s_A, s_B) \;\Bigm|\;&
s_A, s_B \in \mathcal{S}(x)\cup\{S^{\text{GT}}\}, \\
& s_A \neq s_B
\Bigr\}.
\end{aligned}
\end{equation}

To obtain preference annotations, we employ Gemini-2.0-Flash to generate multi-dimensional scores and explanations for each candidate in a pair, covering
\emph{Text Fidelity \& Intelligibility, Speaker Similarity,
Prosody \& Expressiveness, and Naturalness \& Audio Quality},
followed by a final binary decision indicating which speech is better.
For each speech sample, Gemini assigns a score in $[0,10]$ for each dimension, and the total score is computed by summation.
The final preference label is determined by comparing the total scores:
\begin{equation}
    \ell_{better} =
    \begin{cases}
    speech A, & \text{if } T_A > T_B,\\[2pt]
    speech B, & \text{otherwise}.
    \end{cases}
\end{equation}
This process yields a high-quality dataset of speech preference pairs for speech reward modeling. In \textit{Appendix}~\ref{app:data_control}, we present more details about how to remove nosiy samples and human verification criteria to improve the quality of dataset.

\begin{figure*}[t]
    \centering
    % \vspace{-5mm}
    \includegraphics[width=1.0\textwidth]{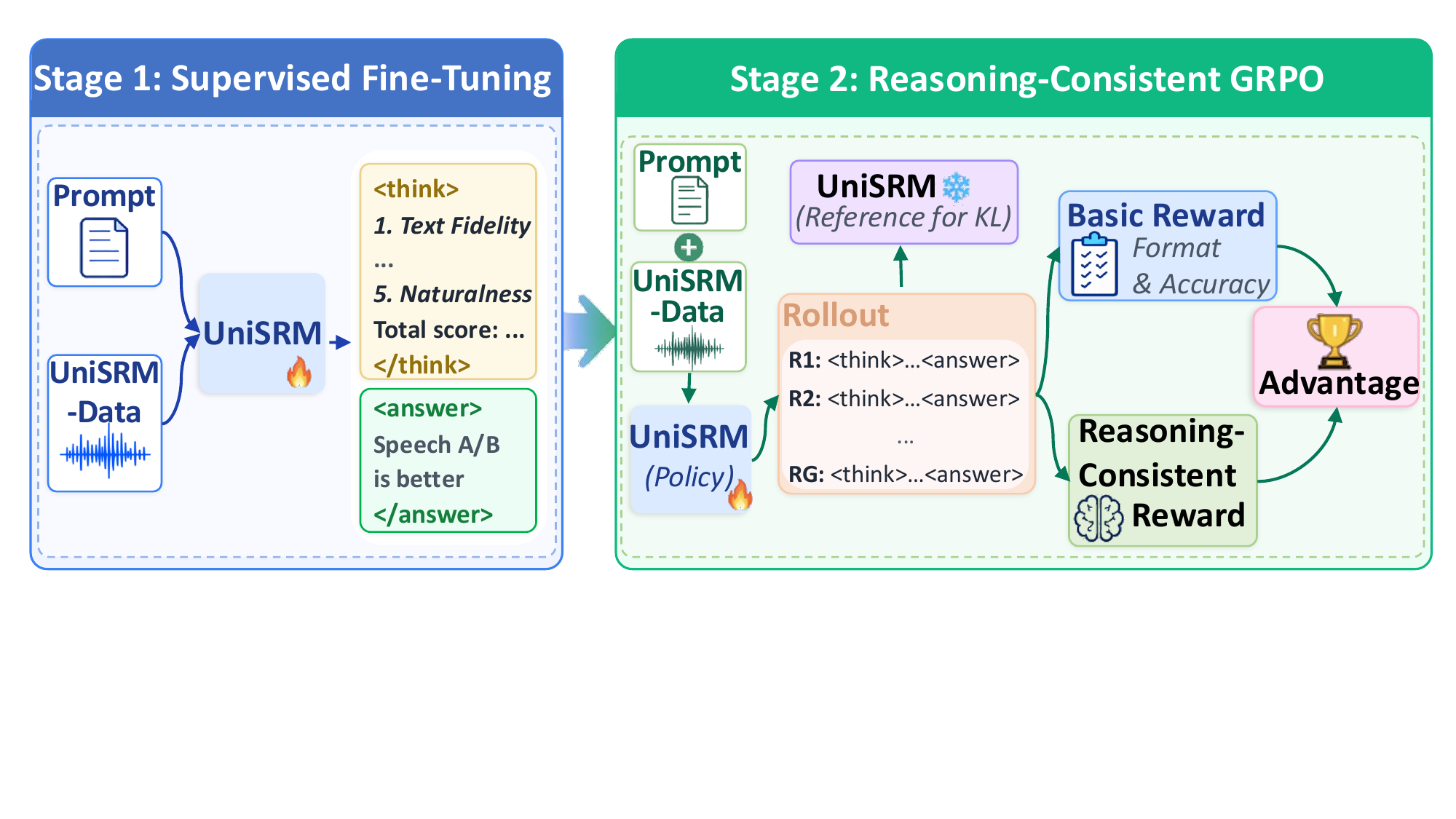}
    \vspace{-85pt}
    \caption{Our proposed two-stage framework of UniSRM.} 
    \label{fig:main}
    \vspace{-15pt}
\end{figure*}

\paragraph{Task 2: Utterance-Level Speech Quality Assessment.}
For single-sample speech quality assessment, we directly leverage the public QualiSpeech~\cite{wang2025qualispeech} dataset,
which provides MOS-like annotations across seven perceptual aspects: \emph{Noise, Distortion, Speed, Continuity, Naturalness, Listening effort, and Overall quality} (each in $[1,5]$).
To align with our reasoning-based training format, we treat QualiSpeech's textual rationales as the model's intermediate reasoning.
Specifically, \emph{Noise description, Distortion description, Unnatural pause, Feeling of voice, together with a
Natural language description} are used as the reasoning trace,
while the seven aspect scores are used as the final structured answer.
This design enables UniSRM to learn fine-grained, interpretable quality assessment rather than a black-box scalar prediction.

\paragraph{Task 3: Scenario-Aware Style Coherency conditioned on Textual Context.}

Speech quality cannot always be judged from an isolated utterance: in expressive narration or voice acting, speech should match the scene context and target emotion. We therefore define scenario-aware style consistency evaluation, where a model compares two speech samples conditioned on a textual scenario and paragraph context.

Using ESD \citep{zhou2022esd}, we treat each original recording as the positive sample, and use GPT-4.1 \citep{openai2024gpt4technicalreport} to generate a coherent scenario description and paragraph context conditioned on the utterance text and emotion label. Hard negatives are created by sampling one corruption type per instance: (i) real-speech negatives from ESD with mismatched text and/or emotion, or (ii) TTS negatives synthesized by open-source TTS systems, optionally with GPT-4.1–generated mismatching text under the same scenario. 

For TTS negatives, we assign prompt speech from ESD or LibriTTS-R~\cite{koizumi2023librittsr}, enforce emotion mismatch, and randomize controllable attributes (e.g., role, speaking rate, style) while filtering accidental emotion matches. Finally, given the scenario context, we use Gemini-2.5-Pro to provide multi-dimensional scores and rationales along \emph{Text Fidelity, Scenario Style Match, Naturalness/Audio Quality}. 
In this way, we generate a bilingual dataset for both Chinese and English.
\paragraph{Task 4: Multi-Turn Dialogue Speech Evaluation conditioned on Dialogue History.}

Existing dialogue speech reward modeling is often restricted to single-turn settings. We introduce a multi-turn dialogue speech evaluation task conditioned on spoken dialogue-history, enabling reward modeling of dialogue-level consistency.

Based on DailyTalk \citep{lee2022dailytalk}, for each dialogue, we randomly sample a target turn $t$ and treat the speech of turns ${1,\dots,t-1}$ as the dialogue history. The original speech at turn $t$ is used as a positive candidate.
To reduce shortcut learning from recording conditions (e.g., \textit{real vs.\ synthesized}), we also synthesize a subset of positives using the open-source TTS systems as negatives.

We construct hard negatives along two axes: text mismatch and audio mismatch. We randomly sample from three categories: (i) text-only negatives generated by GPT-4.1 conditioned on the multi-turn history (e.g., intent/consistency errors), (ii) audio-only negatives that keep the text fixed but alter speaker/prosody/emotion (e.g., speaker swap or prompt speaker mismatch), and (iii) mixed negatives that combine both. All candidates are synthesized with multiple open-source TTS systems to increase diversity.
Given the dialogue-history and two candidate samples, we use Gemini-2.5-Pro to provide multi-dimensional scores and rationales along \emph{Intent Matching, Speaker Consistency, Contextual Consistency, Emotion, Naturalness}.

In conclusion, we partition \textsc{UniSRM-Data} into three disjoint subsets: supervised fine-tuning $\mathcal{D}_{\text{SFT}}$, RL training $\mathcal{D}_{\text{GRPO}}$, and a test set \textsc{UniSRM-Bench}. 
To ensure high label reliability for both optimization and evaluation, we additionally perform human verification on \emph{both} $\mathcal{D}_{\text{RL}}$ and \textsc{UniSRM-Bench}, retaining only samples that match the majority-vote human preference.
Detailed human verification criteria is provided in Appendix~\ref{app:human_criteria}.
We summarize the task-specific evaluation dimensions in Appendix~\ref{app:task_dimensions} (Table~\ref{tab:task_dimensions}).

%% file: content/03method.tex
\section{Method}

In this section, we present the complete training pipeline of UniSRM, as illustrated in Figure~\ref{fig:main}.
Initially, we utilize SFT to adapt the model to diverse evaluation tasks and standardize the output scoring format. 
Subsequently, to further align the model with human preferences and encourage diversity in the reasoning process, we employ RL on a manually curated high-quality dataset $\mathcal{D}_{\text{RL}}$.

\subsection{Supervised Fine-Tuning (SFT)}
\label{sec:sft}
Speech understanding models inherently can evaluate generated speech from multiple perspectives with rational and interpretable judgments, making them well-suited for speech reward modeling.
Therefore, we adopt Qwen2.5-Omni-7B-thinker~\cite{xu2025qwen2} as the backbone and modify its system prompts (Appendix~\ref{appendix:prompt}) to enforce a deterministic and structured output format.
As shown in Stage 1 of Figure~\ref{fig:main}, we train a multi-task speech reward model that supports multi-dimensional reasoning based on $\mathcal{D}_{\text{SFT}}$.
This design makes the model produce \emph{dimension-wise evidence} before outputting the final preference/score,
which is critical for interpretability and provides a stable foundation for subsequent RL optimization.

Although our data cover four tasks with different inputs, they can be unified as a conditional generation problem.
Each training instance is represented as $(x, o)$, where $x$ is the task-specific input prompt, e.g., text content and audio clips, and $o$ is the target structured output.
Given an input $x$, UniSRM outputs a two-part response:
\vspace{-5pt}
\begin{equation}
\begin{aligned}
o &= \pi_\theta(x) \\
  &= \verb|<think>|\,\hat{r}\,\verb|</think><answer>|\,\hat{y}\,\verb|</answer>|.
\end{aligned}
\label{eq:structured_output}
\end{equation}
Here, $\hat{r}$ is an explicit reasoning trace containing \emph{dimension-wise} scores and short explanations, e.g., text fidelity, speaker similarity and so on.
The final output $\hat{y}$ is task-dependent: it is a binary preference decision for pairwise tasks (Task~1/3/4),
or a MOS-like structured score for the pointwise quality task (Task~2).
We fine-tune the model using standard autoregressive maximum likelihood on the full target sequence $o$:
\vspace{-10pt}
\begin{equation}
\label{eq_sft}
\mathcal{L}_{\text{SFT}}(\theta)
= -\mathbb{E}_{(x,o)\sim \mathcal{D}_{\text{SFT}}}
\sum_{t=1}^{|o|}
\log \pi_\theta(o_t \mid o_{<t}, x).
\end{equation}
After SFT, the resulting policy $\pi_\theta$ serves as a stable initialization for the GRPO stage,
where we further improve the correctness, diversity and reliability of the reasoning process.

\subsection{Reinforcement Learning with GRPO}
\label{sec:grpo}
\input{tables/all_task_main}

\input{tables/all_task_abla}

\input{tables/task1_abla}

\input{tables/task2_abla}

\input{tables/task3_en_abla}

\input{tables/task4_abla}

While SFT teaches UniSRM to imitate the judge-generated rationales and final decisions, it does not explicitly
optimize reward-aligned correctness, and may lead the model to learn fixed pattern reasoning.
Therefore, we further optimize UniSRM with Group Relative Policy Optimization (GRPO)
to improve reasoning diversity and reliability.

As shown in Figure~\ref{fig:main}, for each training prompt $x$, we sample $G$ responses
from the current policy $\pi_\theta$:
\vspace{-5pt}
\begin{equation}
o^{(g)} \sim \pi_\theta(\cdot \mid x), \quad g=1,\dots,G,
\end{equation}
where each output follows our structured format (Eq.~\ref{eq:structured_output}).
We combine three complementary reward components:
\vspace{-5pt}
\begin{equation}
\label{eq:total_reward}
R(x,o) = \lambda_{\text{fmt}} R_{\text{fmt}}(o)
       + \lambda_{\text{acc}} R_{\text{acc}}(o) 
       + \lambda_{\text{rc}}  R_{\text{rc}}(o),
\end{equation}
where $R_{\text{fmt}}$ indicates format reward, $R_{\text{acc}}$ optimizes the final answer accuracy,
and $R_{\text{rc}}$ provides reasoning-consistent rewards over each dimension.
For the format reward $R_{\text{fmt}}$, we define $R_{\text{fmt}}(o)\in\{-1,0\}$ as an indicator of whether $o$ matches the output format of different tasks in Appendix~\ref{appendix:prompt}.
If the output violates the required format or fails parsing, we assign a negative reward $-1$ to penalize invalid reasoning traces.
For the accuracy reward $R_{\text{acc}}$ of pairwise tasks (Task~1/3/4), we set
\vspace{-5pt}
\begin{equation}
\label{eq:acc_pairwise}
R_{\text{acc}}(o) =
\mathbf{1}\!\left[y^{(g)} = y^{\star}\right],
\end{equation}
where $y^{\star}\in\{\text{A},\text{B}\}$ is the ground-truth preference label.
For $R_{\text{acc}}$ of Task~2, we use a normalized distance reward on the \emph{overall} score:
\vspace{-5pt}
\begin{equation}
\label{eq:acc_mos}
R_{\text{acc}}(o) = 1 - \frac{\left| \hat{m}_{\text{overall}} - m^{\star}_{\text{overall}} \right|}{m_{\max}-m_{\min}},
\end{equation}
and clamp it to $[0,1]$, where $(m_{\min},m_{\max})=(1,5)$ by default, $\hat{m}_{\text{overall}}$ and $m^{\star}_{\text{overall}}$ denote predicted and ground-truth overall score.

\paragraph{Reasoning-Consistent Rewards (RCR-GRPO).}
A key challenge of RL on judge-style rationales is that optimizing only the final answer may encourage shallow or inconsistent reasoning, e.g., a correct label but mismatched reasoning.
To address this, we introduce Reasoning-Consistent Rewards $R_{\text{rc}}$ to directly supervise the dimension-wise scoring behavior inside \texttt{<think>}.

For pairwise tasks (Task~1/3/4), each output contains dimension-wise scores for both candidates, e.g.,
$\mathbf{a}=[a_1,\dots,a_D]$ for Speech~A and $\mathbf{b}=[b_1,\dots,b_D]$ for Speech~B.
$D=4,3,5$ is the number of dimensions for task~1/3/4, respectively.
We compute a \emph{dimension-wise preference consistency} reward:
\begin{equation}
\label{eq:rc_pairwise}
\begin{aligned}
R_{\text{rc}}(o)
= \frac{1}{D}\sum_{i=1}^D
\mathbf{1}\!\Bigl[
&\text{sign}(a_i-b_i) \\
&= \text{sign}(a_i^{\star}-b_i^{\star})
\Bigr].
\end{aligned}
\end{equation}
where $(\mathbf{a}^{\star},\mathbf{b}^{\star})$ are the ground-truth dimension scores,
and $\text{sign}(\cdot)\in\{-1,0,+1\}$.
Intuitively, $R_{\text{rc}}$ encourages UniSRM to produce aligned per-dimension comparisons rather than only matching the final preference label, which can improve reasoning reliability.

For speech quality assessment (Task~2), the output \texttt{<answer>} provides a $D=7$ aspect score vector
$\hat{\mathbf{m}} \in \{1,\dots,5\}^D$.
We compute a normalized reward:
\begin{equation}
\label{eq:rc_mos}
R_{\text{rc}}(o)
= 1 - \frac{1}{D}\sum_{k=1}^D \frac{|\hat{m}_k - m_k^{\star}|}{m_{\max}-m_{\min}},
\end{equation}
and clamp it to $[0,1]$.
This reasoning-consistent rewards provide targeted supervision over the intermediate reasoning process,
improving diversity while reducing inconsistencies between generated rationales and final decisions.

\paragraph{Group-wise advantage normalization.}
As shown in Equation~\ref{eq:adv}, we compute advantages by normalizing rewards within each prompt group, where $\mu(x)$ and $\sigma(x)$ denote the mean and standard deviation of $\{R^{(g)}\}_{g=1}^G$.
This relative normalization reduces reward scale sensitivity and encourages meaningful comparisons across rollouts.
\begin{equation}
\label{eq:adv}
A^{(g)} = \frac{R^{(g)} - \mu(x)}{\sigma(x)+\epsilon}.
\end{equation}

\paragraph{GRPO objective with KL regularization.}
Let $\pi_{\theta_{\text{old}}}$ denote the policy used to generate rollouts.
GRPO optimizes a clipped policy gradient objective:
\begin{equation}
\label{eq:grpo_obj}
\begin{aligned}
\mathcal{J}(\theta)
= \mathbb{E}_{x}\,\mathbb{E}_{g=1}^G 
\Bigl[
\min\!\Bigl( 
\rho_\theta^{(g)} A^{(g)}, &
\text{clip}(\rho_\theta^{(g)},\\1-\epsilon,1+\epsilon)\, A^{(g)}
\Bigr)
\Bigr].
\end{aligned}
\end{equation}
where $\rho_\theta^{(g)}=\frac{\pi_\theta(o^{(g)}\mid x)}{\pi_{\theta_{\text{old}}}(o^{(g)}\mid x)}$.
To prevent excessive drift from the supervised initialization, we add a KL penalty against a reference policy
$\pi_{\text{ref}}$ (the SFT model checkpoint):
\begin{equation}
\label{eq:grpo_loss}
\begin{aligned}
\mathcal{L}_{\text{GRPO}}(\theta)
= -\mathcal{J}(\theta)
+ &\\
\beta \cdot \mathbb{E}_{x}\,\mathbb{E}_{g=1}^G
\Bigl[
\mathrm{KL}\!\Bigl(&
\pi_\theta(\cdot\mid x)\,\|\,\pi_{\text{ref}}(\cdot\mid x)
\Bigr)
\Bigr].
\end{aligned}
\end{equation}

%% file: tables/all_task_main.tex
\begin{table*}[tb]
\centering
% \small
\setlength{\tabcolsep}{1.5mm}
\begin{tabular}{cccccc}
\toprule
\multirow{2}{*}{\textbf{Model}}            &
\multirow{1}{*}{\textbf{T1}}    &
\multirow{1}{*}{\textbf{T2}} &
\multirow{1}{*}{\textbf{T3-En}}    &
\multirow{1}{*}{\textbf{T3-Zh}}    &
\multirow{1}{*}{\textbf{T4}}   
\\
& $acc\uparrow$ & $acc\uparrow/pcc\uparrow$ & $acc\uparrow$ & $acc\uparrow$ & $acc\uparrow$
\\  \midrule
% \multicolumn{4}{c}{\textbf{Objective Evaluation}} 
\textit{\textbf{Objective Metrics}}
\\ [2pt]
WER & 59.24 &-/- & 61.44 & 56.92 & 84.10\\
SIM & 47.99 &-/- & - & - & - \\
UTMOS & 50.20 & -/0.449 & 33.21 & 48.19 & 40.48 \\
DNSMOS & 49.80& -/0.274 &53.51 &63.04 & 50.79 \\
\hline
% \textbf{Subjective Evaluation}
% \\ [2pt]
% MOS \\
% \hline
\textit{\textbf{Proprietary Models}} \\ [2pt]
% \hdashline
% \textbf{Zero-shot Prompting} \\
GPT-4o-Audio~\citep{hurst2024gpt4o}  & 61.04 & 24.60/0.060 &64.02 & 64.82 & 71.96\\
Gemini-2.5-Flash &60.44 & 34.50/0.522&65.68&71.74&71.43\\
Gemini-2.5-Pro &60.67 & 28.93/0.517 &67.31 & 63.47 & 82.40\\
\hline
% \hdashline
\textit{\textbf{Open-Source Models}} \\
Kimi-Audio-7B~\citep{ding2025kimi} & 52.81 & 22.93/0.209 & 71.22 & 69.70&64.29\\
MiMo-Audio-7B~\citep{coreteam2025mimoaudio} & 50.40 & 26.36/0.158 &47.97&42.49&59.52 \\
Qwen2.5-Omni-7B~\cite{xu2025qwen2} & 51.20 & 24.03/0.289 & 49.45&52.17& 56.35  \\

SpeechJudge~\cite{zhang2025speechjudge} & 57.20 &-/-&-&-&-\\

\hline
\textit{\textbf{Proposed Method}} \\ [2pt]

UniSRM(Ours) & 65.06 & 39.74/0.551 & 85.61 & 91.30 & 88.89 \\
\bottomrule
\end{tabular}

\caption{Overall results on \textsc{UniSRM-Bench}. 
\textbf{T1}: utterance-level pairwise preference judgement. 
\textbf{T2}: fine-grained speech quality scoring. 
\textbf{T3-En} / \textbf{T3-Zh}: scenario-aware style consistency preference in English and Chinese. 
\textbf{T4}: multi-turn dialogue speech evaluation conditioned on spoken dialogue history.}
\vspace{-10pt}
 
\label{tab:overall}
\end{table*}

%% file: tables/all_task_abla.tex
\begin{table}[tb]
\centering
\setlength{\tabcolsep}{0.25mm}
\begin{tabular}{cccccc}
\toprule
\multirow{1}{*}{\textbf{Model}}            &
\multirow{1}{*}{\textbf{T1}}    &
\multirow{1}{*}{\textbf{T2}}&
\multirow{1}{*}{\textbf{T3-En}}    &
\multirow{1}{*}{\textbf{T3-Zh}}    &
\multirow{1}{*}{\textbf{T4}}    
% & \multirow{1}{*}{\textbf{Format-Acc}}
% & \multirow{1}{*}{\textbf{Rank Correlation}}
\\  \midrule
UniSRM(Ours) & 65.06 & 39.74&85.61 &91.30 & 88.89 \\
\hline
$w/o$ RCR-GRPO &60.44 &37.58& 80.81 & 81.42 & 82.54  \\
$w/o$ GRPO &  60.24&39.20 & 67.16&70.95 & 74.60 \\
\bottomrule
\end{tabular}
\caption{Ablation results over all tasks. Results are reported as accuracy ($\uparrow$, \%).}
\vspace{-12pt}
\label{tab:ablation_rc}
\end{table}

%% file: tables/task1_abla.tex
\begin{table*}[tb]
\centering
\setlength{\tabcolsep}{1.0mm}
\begin{tabular}{cccccc}
\toprule
\multirow{1}{*}{\textbf{Model}}            &
\multirow{1}{*}{\textbf{Text}} &
\multirow{1}{*}{\textbf{Sim}} &
\multirow{1}{*}{\textbf{Expressiveness}} &
\multirow{1}{*}{\textbf{Naturalness}} &
\multirow{1}{*}{\textbf{AVG}} 
\\  \midrule
UniSRM(Ours)  & 83.33&	62.25&	61.24&	43.98&	62.70\\
$w/o$ RCR-GRPO & 76.89&	59.22&	60.23&	39.76&	59.03\\
$w/o$ GRPO & 83.53&	57.83&	59.84&	42.37&	60.89 \\
\bottomrule
\end{tabular}
\caption{Multi-dimensional results of pair-wise speech preference task. Results are reported as accuracy ($\uparrow$, \%).}
\vspace{-6pt}
\label{exp_task1_abla}
\end{table*}

%% file: tables/task2_abla.tex
\begin{table*}[tb]
\centering
\setlength{\tabcolsep}{0.5mm}
\begin{tabular}{ccccccccc}
\toprule
\multirow{1}{*}{\textbf{Model}}            &
\multirow{1}{*}{\textbf{Noise}} &
\multirow{1}{*}{\textbf{Distortion}} &
\multirow{1}{*}{\textbf{Speed}} &
\multirow{1}{*}{\textbf{Continuity}} &
\multirow{1}{*}{\textbf{Effort}} &
\multirow{1}{*}{\textbf{Naturalness}}&
\multirow{1}{*}{\textbf{Overall}} &
\multirow{1}{*}{\textbf{AVG}}
\\  \midrule
QualiSpeech  & 0.686& 0.518 &\underline{0.250} &0.459& 0.475& \underline{0.486} &\textbf{0.572} & 0.492\\
\hline
\textit{\textbf{UniSRM}} \\
UniSRM(Ours) & \textbf{0.754}	&\textbf{0.547} &	0.209&	\textbf{0.526}	&	\underline{0.478}&0.473&	\underline{0.551} & \textbf{0.505}\\
$w/o$ RCR-GRPO & 0.688&	\underline{0.528}&	0.233&	\underline{0.512}&	0.446 &	0.418	&0.542 & 0.481 \\
$w/o$ GRPO & \underline{0.714}	&0.514	&\textbf{0.268}	&0.471	&\textbf{0.481} & \textbf{0.506} &	0.534 & \underline{0.498} \\

\bottomrule
\end{tabular}
\caption{Multi-dimensional results on the QualiSpeech dataset. Results are reported as PCC$(\uparrow)$.}
\vspace{-10pt}
\label{exp_task4_abla}
\end{table*}

%% file: tables/task3_en_abla.tex
\begin{table*}[t]
\centering
% \small
% \setlength{\tabcolsep}{6pt}
\setlength{\tabcolsep}{0.45mm}
\begin{tabular}{ccccccccc}
\toprule
\multirow{2}{*}{\textbf{Model}} &
\multicolumn{4}{c}{\textbf{English}} &
\multicolumn{4}{c}{\textbf{Chinese}} \\
\cmidrule(lr){2-5}\cmidrule(lr){6-9}
& \textbf{Text} & \textbf{Scenario} & \textbf{Naturalness} & \textbf{AVG}
& \textbf{Text} & \textbf{Scenario} & \textbf{Naturalness} & \textbf{Avg} \\
\midrule
UniSRM(Ours)   & 87.45 & 85.61 & 81.00 & 84.69 & 86.76 & 91.11 & 88.14 & 88.67 \\
w/o RCR-GRPO   & 89.30 & 80.81 & 76.20 & 82.10 & 84.78 & 81.03 & 78.66 & 81.49 \\
w/o GRPO       & 87.64 & 67.16 & 63.84 & 72.88 & 85.38 & 70.75 & 68.97 & 75.03 \\
\bottomrule
\end{tabular}
\caption{Multi-dimensional results on the scenario-aware speech preference task. Results are reported as accuracy ($\uparrow$, \%).}
\vspace{-10pt}
\label{exp_task2_en_zh_abla}
\end{table*}

%% file: tables/task4_abla.tex
\begin{table*}[tb]
\centering
\setlength{\tabcolsep}{1.0mm}
\begin{tabular}{ccccccc}
\toprule
\multirow{1}{*}{\textbf{Model}}            &
\multirow{1}{*}{\textbf{Intent}} &
\multirow{1}{*}{\textbf{Sim}} &
\multirow{1}{*}{\textbf{Context}} &
\multirow{1}{*}{\textbf{Emotion}} &
\multirow{1}{*}{\textbf{Naturalness}} &
\multirow{1}{*}{\textbf{Avg}} 
\\  \midrule

UniSRM(Ours) & 86.51&	72.22&	83.33&	88.89&	88.89&	83.97\\
$w/o$ RCR-GRPO & 68.25&	57.14&	67.46&	65.87&	68.25&	65.39 \\
$w/o$ GRPO & 69.15&	53.17&	69.05&	50.79&	61.90&	60.81\\

\bottomrule
\end{tabular}
\caption{Multi-dimensional results on the dialogue preference task. Results are reported as accuracy ($\uparrow$, \%).}
\vspace{-15pt}
\label{exp_task3_abla}
\end{table*}

%% file: content/04exp.tex
\section{Experiments and Analyses}

\subsection{Evaluation Metrics}
% \subsubsection{Objective Evaluation}
We evaluate UniSRM using two primary metrics: Accuracy (ACC) for pairwise decision tasks and Pearson Correlation Coefficient (PCC) for score-based assessment.
For the preference-based tasks (Task~1, Task~3, and Task~4), the model outputs a binary decision
$\hat{y}\in\{\text{Speech A}, \text{Speech B}\}$ in the \texttt{<answer>} tag.
We report ACC as the proportion of samples whose predicted preference matches the ground-truth label.
For the fine-grained speech quality task (Task~2), the model predicts MOS-like scores for multiple aspects.
Following the standard practice in QualiSpeech\citep{wang2025qualispeech}, we use PCC to measure how well the predicted scores correlate with
the ground-truth scores.

\subsection{Main Results} \label{sec:results_overall}

Table~\ref{tab:overall} presents the main results on \textsc{UniSRM-Bench}. We compare four groups of approaches: objective metrics, proprietary AudioLLM models, open-source AudioLLM models, and our proposed UniSRM. 

Across all tasks, UniSRM attains the best overall performance.
Notably, UniSRM exhibits a larger advantage on context-dependent evaluations (T3/T4), where the judge must integrate textual context or multi-turn conversational context rather than relying solely on local acoustic cues. 
This observation further suggests that, under rich textual or spoken contextual conditions, many existing models still struggle to deliver reliable judgments.
For the MOS-style assessment in T2, UniSRM shows the most consistent alignment with annotated quality scores, providing additional evidence for the effectiveness of our proposed training and optimization strategy.

\subsection{Ablations}
\label{sec:ablation_rc}

Table~\ref{tab:ablation_rc} analyzes the effect of our training methods. 
We include two key variants: \textit{w/o GRPO}, which removes reinforcement learning and keeps only SFT; and \textit{w/o RCR-GRPO}, which still applies GRPO but uses only an accuracy-based reward that measures the correctness of the final decision, without our reasoning-consistent rewards (RCR-GRPO).

Overall, incorporating GRPO consistently improves performance over SFT-only (\textit{w/o GRPO}) training, indicating that on-policy optimization better aligns the judge with evaluation objectives beyond supervised imitation. More importantly, comparing \textit{w/o RCR-GRPO} against our method shows that adding RCR further improves results across tasks. This suggests that optimizing only the final accuracy is insufficient: without RCR-GRPO, the model may obtain high outcome reward on some samples via shortcut behaviors, while producing rationales that are weakly grounded in the provided context or even inconsistent with the final choice, which in turn degrades overall reliability and performance.

\subsection{Fine-grained Analysis Across Tasks and Dimensions}
\label{sec:fine_grained}

% Tables~\ref{exp_task1_abla}, \ref{exp_task4_abla}, \ref{exp_task2_en_zh_abla}, and \ref{exp_task3_abla} 
Tables~\ref{exp_task1_abla}--\ref{exp_task3_abla}
report a dimension-level breakdown for all four tasks, offering a closer look at \emph{where} the improvements come from and \emph{how} Reasoning-Consistent Rewards (RCR) shape the reasoning behavior. Overall, UniSRM demonstrates strong and well-balanced performance across nearly all dimensions, rather than improving a single metric at the expense of others.

\paragraph{Task 1.}
UniSRM improves the preference reliability by consistently strengthening multiple complementary factors, including text fidelity, speaker similarity, expressiveness, and naturalness. Importantly, compared to \textit{w/o RCR-GRPO}, our UniSRM shows more stable gains on the harder perceptual dimensions (e.g., naturalness), which indicates RCR helps maintain a more correct comparison across dimensions.

\paragraph{Task 2.}
For MOS-style assessment, UniSRM achieves the most consistent alignment with human-annotated aspect scores across the seven criteria. 
This indicates that RCR is not only beneficial for pairwise judgments but also improves calibration for multi-aspect scoring. 
In contrast, \textit{w/o RCR-GRPO} tends to underperform across several aspects, suggesting that optimizing only an outcome-based reward provides insufficient guidance for fine-grained scoring behavior and may distort the internal assessment criteria.

\paragraph{Task 3.}
Table~\ref{exp_task2_en_zh_abla} further highlights the advantage of RCR in context-dependent evaluation. UniSRM with RCR yields consistently better performance on scenario-related dimensions in both English and Chinese.

\paragraph{Task 4.}
For dialogue-conditioned judging, UniSRM also shows robust improvements across different aspects.
These dimensions are tightly coupled and require long-context integration over the spoken dialogue history. 
Our UniSRM provides the largest benefit by encouraging dimension-wise assessments to remain consistent with the final decision.

\paragraph{When accuracy-only GRPO can be worse than SFT.}
A noteworthy finding across Tables~\ref{exp_task1_abla}--\ref{exp_task3_abla} is that \textit{w/o RCR-GRPO} (Accuracy-only GRPO) sometimes underperform \textit{w/o GRPO} (SFT-only) on certain dimensions. 
This phenomenon suggests that optimizing with only an accuracy-based reward may cause the model's reasoning to drift: the policy can exploit shortcuts to obtain favorable outcome rewards while degrading dimension-level judgments, which ultimately hurts overall reliability. 
Our RCR-GRPO mitigates this issue by directly supervising dimension-wise consistency, leading to more stable improvements across tasks and dimensions.

\begin{table*}[t]
\centering
\begin{tabular}{ccccccc}
\toprule
\multirow{2}{*}{\textbf{Model}} & \multicolumn{2}{c}{\textbf{BVCC}} & \multicolumn{2}{c}{\textbf{SOMOS-Clean}} & \multicolumn{2}{c}{\textbf{SOMOS-Full}} \\
\cmidrule(lr){2-3}\cmidrule(lr){4-5}\cmidrule(lr){6-7}
 & $pcc\uparrow$ & $acc\uparrow$ & $pcc\uparrow$ & $acc\uparrow$ & $pcc\uparrow$ & $acc\uparrow$ \\
\midrule
DNSMOS              & 0.2990 & --      & 0.0479 & --      & 0.0528 & -- \\
Qwen2.5-Omni-7B     & 0.2563 & 25.57   & 0.1561 & 23.17   & 0.1484 & 22.70 \\
Gemini-2.5-Flash    & 0.3420 & 29.84   & 0.2498 & 29.06   & 0.2156 & 27.83 \\
Gemini-2.5-Pro      & 0.3390 & 27.42   & 0.2009 & 30.71   & 0.2218 & 33.94 \\
\textbf{UniSRM}     & \textbf{0.4977} & \textbf{49.16} & \textbf{0.2612} & \textbf{41.70} & \textbf{0.2347} & \textbf{52.97} \\
\bottomrule
\end{tabular}
\caption{Cross-dataset generalization on speech quality datasets.}
\label{tab:cross_dataset}
\end{table*}

\subsection{Cross-dataset generalization on speech quality datasets}
To validate the generalization ability of UniSRM, we further evaluate it on external human-labeled speech quality datasets, including BVCC~\cite{cooper2021voices} and SOMOS~\cite{maniati2022somos}, where SOMOS is entirely unseen during training.
Following prior MOS prediction settings, we report both PCC with the human MOS scores and ACC after discretizing MOS into integer bins. 
For SOMOS, whose ground-truth MOS is fractional, we keep the original decimal scores for PCC and round them only when computing ACC. 

As shown in Table~\ref{tab:cross_dataset}, these results demonstrate that UniSRM generalizes well beyond the training distribution and is competitive with, or better than, Gemini-family baselines on these human-annotated benchmarks. This finding suggests that UniSRM is not simply overfitting LLM-generated labels, but instead learns reward signals that transfer effectively to unseen domains under human supervision.

%% file: content/05_conclusion.tex
\section{Conclusion}
In this work, we aim to develop a unified speech reward model that can provide multi-dimensional, interpretable judgments for speech evaluation.
To this end, we introduce \textsc{UniSRM-Data} and \textsc{UniSRM-Bench}, which jointly cover speech evaluation tasks ranging from \emph{utterance-level quality} to \emph{context-level coherency}.
Building on this unified data formulation, we present UniSRM with a two-stage pipeline, enabling a single model to support both pairwise preference decisions and fine-grained scoring with explicit reasoning traces.
Furthermore, we propose RCR-GRPO to directly supervise dimension-wise reasoning during RL, improving the reliability of rationales.
Experimental results on \textsc{UniSRM-Bench} demonstrate that UniSRM yields more accurate, human-aligned, and robust judgments across all tasks.

\section*{Ethics Statement}
All models and datasets used in this paper are employed in compliance with their ethical guidelines and licensing terms. When we synthesize speech using open-source TTS systems, we also use these models under their respective licenses. We have provided a clear description of data sources and will release UniSRM-Data and UniSRM-Bench under a suitable open license for research use.
Human annotations are conducted by annotators who receive clear instructions, qualification checks, and fair compensation upon completion. We do not collect personally identifiable information during the annotation process. Following common practice, we employ LLMs for semi-automated annotation assistance and model evaluation to improve efficiency and consistency.

\section*{Limitations}

Despite promising results, our work has several limitations.
For example, our current benchmark coverage is limited for challenging scenarios such as heavy accents and overlapped speech.
Extending \textsc{UniSRM-Data} and \textsc{UniSRM-Bench} to broader acoustic conditions and application settings
remains an important direction for future work.
Moreover, training and inference with speech-LLM backbones, multi-sample rollouts, and GRPO-based optimization incur non-trivial computational cost.
This may limit scalability to larger backbones, larger rollout size $G$, or broader datasets, and may hinder low-latency deployment
when \textsc{UniSRM} is used as an online speech judge.
Future work could explore more efficient architectures, lightweight distillation, and caching strategies to reduce inference overhead.

Despite these limitations, we believe UniSRM provides a valuable and practical foundation for the community:
an interpretable, multi-dimensional, and unified speech reward modeling framework, together with a comprehensive dataset and benchmark,
which can facilitate more reliable reward-based evaluation and optimization for speech generation systems.

\section*{Acknowledgement}
This work is partially supported by the General Research Fund from the Research Grants Council of Hong Kong SAR Government (Project No. 14202623) and National Natural Science Foundation of China (62076144).

%% file: appendix/system_prompt.tex
\section{Prompt for UniSRM}
\label{appendix:prompt}
This section presents the detailed prompts used by UniSRM across all tasks.
As illustrated in Figures~\ref{prompt_task1}, \ref{prompt_qualispeech}, \ref{prompt_task3_en}, and \ref{prompt_dialogue},
we provide the complete system prompt templates for Task~1--4, respectively.
Figure~\ref{prompt_for_scene_generation} shows the prompt used to query GPT-4.1 for constructing the scenario and textual context in Task~3.

\begin{center}
\begin{tcolorbox}[
    breakable,
    colback=lavenderback, 
    colframe=lavenderframe,
    arc=1mm,
    boxrule=0.8pt,
    left=5pt, right=5pt, top=5pt, bottom=5pt,
    fonttitle=\bfseries,
    title=Prompt for Speech Evaluation
]
\small
\setlist[description]{leftmargin=1em, style=nextline} % 控制缩进

You are an expert speech evaluator.

\smallskip
\textbf{Inputs:}
\begin{description}[nosep, leftmargin=1.5em]
    \item[{[Reference Text]}] only for checking text fidelity.
    \item[{[Prompt Speech]}] only for speaker identity similarity.
    \item[{[Speech A, B]}] candidate speech to evaluate.
\end{description}

\smallskip
\textbf{Your job:}
\begin{enumerate}[leftmargin=1.5em, nosep]
    \item Score Speech A and B on FOUR dimensions (0--10 each):
    \begin{itemize}[leftmargin=1em, nosep, label=\small$\circ$]
        \item (1) Text Fidelity \& Intelligibility
        \item (2) Speaker Similarity to Prompt Speech
        \item (3) Prosody \& Expressiveness
        \item (4) Naturalness \& Audio Quality
    \end{itemize}
    \item For Speaker Similarity, use ONLY voice cues (timbre, pitch, accent, style, etc.), not text content.
    \item Compute Total\_A and Total\_B (no ties allowed).
    \item Decide which speech is better overall.
\end{enumerate}

\smallskip
\textbf{Hard constraints:}
\begin{itemize}[leftmargin=1.5em, nosep]
    \item In \texttt{<think>}: include scores, explanations, and a [Comparison summary] (2--4 sentences).
    \item In \texttt{<answer>}: output EXACTLY \textit{``Speech A is better''} or \textit{``Speech B is better''}.
\end{itemize}

\smallskip
\textbf{Output format:}
\begin{tcolorbox}[colback=codeback, colframe=gray!20, size=minimal, left=5pt, top=2pt, bottom=2pt]
\footnotesize\ttfamily
<think> \\
{[Speech A]} \\
1) Text Fidelity \& Intelligibility: score=a1/10; explanation: ... \\
2) Speaker Similarity to Prompt Speech: score=a2/10; explanation: ...\\
3) Prosody \& Expressiveness Appropriateness: score=a3/10; explanation: ...\\
4) Naturalness \& Audio Quality: score=a4/10; explanation: ...\\
Total\_A = a1+a2+a3+a4 = A\_total \\
{[Speech B]} \\
Similar to Speech A. \\
{[Comparison summary]} \\
- 2–4 sentences explaining the main differences and why the winner is better. \\
</think> \\
<answer>Speech A is better</answer>
\end{tcolorbox}
\end{tcolorbox}

\captionof{figure}{Prompt template used for Task 1 (utterance-level speech A/B preference judgment).}
\label{prompt_task1}
\end{center}

\begin{center}
\begin{tcolorbox}[
    breakable,
    colback=lavenderback, 
    colframe=lavenderframe,
    arc=1mm,
    boxrule=0.8pt,
    left=5pt, right=5pt, top=5pt, bottom=5pt,
    fonttitle=\bfseries,
    title=Prompt for Speech Quality Assessment (QualiSpeech)
]
\small
\setlist[description]{leftmargin=1em, style=nextline}

You are an expert judge for speech quality assessment.

\smallskip
\textbf{Inputs:}
\begin{description}[nosep, leftmargin=1.5em]
    \item[{[Speech]}] one speech sample to evaluate.
\end{description}

\smallskip
\textbf{Aspects (1--5 each; 1=worst, 5=best):}
\begin{enumerate}[leftmargin=1.5em, nosep]
    \item Noise
    \item Distortion
    \item Speed (speaking rate)
    \item Continuity (smoothness / discontinuity)
    \item Naturalness
    \item Listening effort
    \item Overall quality
\end{enumerate}

\smallskip
\textbf{Your job:}
\begin{enumerate}[leftmargin=1.5em, nosep]
    \item Carefully listen to the audio and analyze its quality across all seven aspects.
    \item In \texttt{<think>}, first restate concise aspect descriptions (noise / distortion / unnatural pauses / feeling of voice), then provide a coherent paragraph explaining your overall quality judgment in natural language.
    \item In \texttt{<answer>}, output ONLY the final scores for all seven aspects in a fixed \texttt{key=value} format.
\end{enumerate}

\smallskip
\textbf{Hard constraints:}
\begin{itemize}[leftmargin=1.5em, nosep]
    \item Scores \texttt{N, D, S, C, Na, L, O} MUST be integers in \texttt{[1,5]}.
    \item Use ONLY \texttt{<think>...</think>} and \texttt{<answer>...</answer>}. No extra text.
\end{itemize}

\smallskip
\textbf{Output format (STRICT):}
\begin{tcolorbox}[colback=codeback, colframe=gray!20, size=minimal, left=5pt, top=2pt, bottom=2pt]
\footnotesize\ttfamily
<think> \\
{[Aspect descriptions]} \\
Noise description: ... \\
Distortion description: ... \\
Unnatural pause: ... \\
Feeling of voice: ... \\
\\
{[Natural language description]} \\
A detailed paragraph explaining the perceived quality, covering all aspects. \\
</think> \\
<answer>noise=N; distortion=D; speed=S; continuity=C; naturalness=Na; listening\_effort=L; overall=O;</answer>
\end{tcolorbox}
\end{tcolorbox}

\captionof{figure}{Prompt template used for Task 2 (speech quality assessment with seven MOS-like aspects).}
\label{prompt_qualispeech}
\end{center}

\begin{center}
\begin{tcolorbox}[
    breakable,
    colback=lavenderback, 
    colframe=lavenderframe,
    arc=1mm,
    boxrule=0.8pt,
    left=5pt, right=5pt, top=5pt, bottom=5pt,
    fonttitle=\bfseries,
    title=Prompt for Scenario-Aware Speech Evaluation (EN)
]
\small
\setlist[description]{leftmargin=1em, style=nextline}

You are an expert judge for SCENARIO-AWARE speech evaluation.

\smallskip
\textbf{Inputs:}
\begin{description}[nosep, leftmargin=1.5em]
    \item[{[Scene Context]}] Scenario Description, Paragraph Context, Target Emotion.
    \item[{[Target Text]}] the exact sentence that should be spoken.
    \item[{[Speech A, B]}] two audios for the same target text.
\end{description}

\smallskip
\textbf{Your job:}
\begin{enumerate}[leftmargin=1.5em, nosep]
    \item Evaluate Speech A and Speech B as realizations of the target text under the given context.
    \item Score each speech on THREE dimensions (0--10 each) with 1--2 sentence explanations:
    \begin{itemize}[leftmargin=1em, nosep, label=\small$\circ$]
        \item (1) Text Fidelity \& Intelligibility
        \item (2) Scenario Style Match \hfill \textbf{[CRITICAL]}
        \item (3) Naturalness \& Audio Quality
    \end{itemize}
    \item Compute Total\_A and Total\_B as the sum of the three scores (they MUST be different).
    \item In \texttt{<answer>}, decide which speech is better overall.
\end{enumerate}

\smallskip
\textbf{Dimension hints:}
\begin{itemize}[leftmargin=1.5em, nosep]
    \item \textbf{Text Fidelity \& Intelligibility:} matches the target text; clear and understandable.
    \item \textbf{Scenario Style Match:} emotion and speaking style fit the target emotion and context.
    \item \textbf{Naturalness \& Audio Quality:} human-like, stable, and comfortable to listen to.
\end{itemize}

\smallskip
\textbf{Hard constraints:}
\begin{itemize}[leftmargin=1.5em, nosep]
    \item Output ONLY \texttt{<think>} and \texttt{<answer>}, nothing else.
    \item In \texttt{<think>}: include both [Speech A] and [Speech B] with scores and explanations, and a [Comparison summary] (2--4 sentences).
    \item In \texttt{<answer>}: output EXACTLY \textit{``Speech A is better''} or \textit{``Speech B is better''}.
\end{itemize}

\smallskip
\textbf{Output format:}
\begin{tcolorbox}[colback=codeback, colframe=gray!20, size=minimal, left=5pt, top=2pt, bottom=2pt]
\footnotesize\ttfamily
<think> \\
{[Speech A]} \\
1) Text Fidelity \& Intelligibility: score=a1/10; explanation: ... \\
2) Scenario Style Match: score=a2/10; explanation: ... \\
3) Naturalness \& Audio Quality: score=a3/10; explanation: ... \\
Total\_A = a1+a2+a3 = A\_total \\
{[Speech B]} \\
Similar to Speech A.\\
{[Comparison summary]} \\
- 2--4 sentences highlighting the main differences and why the winner is better. \\
</think> \\
<answer>Speech A is better</answer>
\end{tcolorbox}
\end{tcolorbox}

\captionof{figure}{Prompt template used for Task 3 (Scenario-aware evaluation, EN).}
\label{prompt_task3_en}
\end{center}

\begin{center}
\begin{tcolorbox}[
    breakable,
    colback=lavenderback, 
    colframe=lavenderframe,
    arc=1mm,
    boxrule=0.8pt,
    left=5pt, right=5pt, top=5pt, bottom=5pt,
    fonttitle=\bfseries,
    title=Prompt for Multi-turn Spoken Dialogue Evaluation
]
\small
\setlist[description]{leftmargin=1em, style=nextline}

You are an expert judge for multi-turn SPOKEN dialogues.

\smallskip
\textbf{Inputs:}
\begin{description}[nosep, leftmargin=1.5em]
    \item[{[dialog\_history]}] audios containing all previous turns, up to but NOT including the current turn.
    \item[{[speech\_A]}] candidate response A for the current turn.
    \item[{[speech\_B]}] candidate response B for the current turn.
\end{description}
\smallskip
\textbf{Your job:}
\begin{enumerate}[leftmargin=1.5em, nosep]
    \item Evaluate both candidates A and B as possible next turns given \texttt{dialog\_history}.
    \item Score each candidate on FIVE dimensions (0--10 each) with 1--2 sentence explanations:
    \begin{itemize}[leftmargin=1em, nosep, label=\small$\circ$]
        \item (1) Intent Matching \& Dialogue Act
        \item (2) Speaker Consistency
        \item (3) Contextual Consistency
        \item (4) Emotion \& Prosody Match
        \item (5) Overall Naturalness
    \end{itemize}
    \item Compute the total score for Speech A and Speech B (sum of the five dimensions; totals MUST be different), then decide which speech is better overall.
\end{enumerate}
\smallskip
\textbf{Dimension hints:}
\begin{itemize}[leftmargin=1.5em, nosep]
    \item \textbf{Intent Matching \& Dialogue Act:} does the reply follow the topic and intent appropriately?
    \item \textbf{Speaker Consistency:} does the voice match the same person in relevant turns (timbre, pitch, gender cues, accent, speaking style)?
    \item \textbf{Contextual Consistency:} is the content consistent with previous turns without contradictions?
    \item \textbf{Emotion \& Prosody Match:} is the emotion, tone, and prosody suitable for the current situation?
    \item \textbf{Overall Naturalness:} does it sound like a coherent, natural human reply in this dialogue?
\end{itemize}

\smallskip
\textbf{Hard constraints:}
\begin{itemize}[leftmargin=1.5em, nosep]
    \item Output ONLY \texttt{<think>} and \texttt{<answer>}, nothing else.
    \item In \texttt{<answer>}: output EXACTLY \textit{``Speech A is better''} or \textit{``Speech B is better''}.
\end{itemize}

\smallskip
\textbf{Output format (STRICT):}
\begin{tcolorbox}[colback=codeback, colframe=gray!20, size=minimal, left=5pt, top=2pt, bottom=2pt]
\footnotesize\ttfamily
<think> \\
{[Speech A evaluation]} \\
- Intent Matching \& Dialogue Act: score=a1/10; explanation: ... \\
- Speaker Consistency: score=a2/10; explanation: ... \\
- Contextual Consistency: score=a3/10; explanation: ... \\
- Emotion \& Prosody Match: score=a4/10; explanation: ... \\
- Overall Naturalness: score=a5/10; explanation: ... \\
- Total score for Speech A = a1+a2+a3+a4+a5 = A\_total \\
{[Speech B evaluation]} \\
Similar to Speech A.
 \\
{[Comparison summary]} \\
- Brief comparison of Speech A vs Speech B across the five dimensions (2--4 sentences), explaining why the chosen speech fits the multi-turn context better. \\
</think> \\
<answer>Speech A is better</answer>
\end{tcolorbox}
\end{tcolorbox}

\captionof{figure}{Prompt template used for Task 4 (multi-turn dialogue evaluation).}
\label{prompt_dialogue}
\end{center}

\begin{center}
\begin{tcolorbox}[
    breakable,
    colback=lavenderback, 
    colframe=lavenderframe,
    arc=1mm,
    boxrule=0.8pt,
    left=5pt, right=5pt, top=5pt, bottom=5pt,
    fonttitle=\bfseries,
    title=Prompt for Scenario Context Construction using GPT-4.1.
]
\small
\setlist[description]{leftmargin=1em, style=nextline}

You are a skilled scene and story generator for speech data.

\smallskip
\textbf{Inputs:}
\begin{description}[nosep, leftmargin=1.5em]
    \item[{[Utterance Text]}] a single sentence that will be spoken.
    \item[{[Emotion Label]}] the target emotion of how this sentence is spoken (e.g., Neutral, Angry, Happy, Sad, Surprise).
    \item[{[Target Language]}] the language of the utterance.
\end{description}

\smallskip
\textbf{Your job:}
\begin{enumerate}[leftmargin=1.5em, nosep]
    \item Construct a coherent scenario and short story context where the utterance would naturally appear.
    \item Ensure the scenario and context make the given emotion label reasonable and consistent.
    \item Ensure the context logically leads to the utterance text.
\end{enumerate}

\smallskip
\textbf{Hard constraints:}
\begin{itemize}[leftmargin=1.5em, nosep]
    \item Output MUST be a strict JSON object with exactly two fields: \texttt{scenario\_description} and \texttt{paragraph\_context}.
    \item The language of both fields MUST be \texttt{\{LANG\}}.
    \item Do NOT rewrite or change the utterance text itself.
    \item Make the emotion expression implicitly reasonable; avoid explicitly stating the emotion in every sentence (e.g., do not repeatedly say ``he is angry''), but ensure the situation reflects the target emotion.
\end{itemize}

\smallskip
\textbf{Output format (STRICT JSON):}
\begin{tcolorbox}[colback=codeback, colframe=gray!20, size=minimal, left=5pt, top=2pt, bottom=2pt]
\footnotesize\ttfamily
\{ \\
\quad "scenario\_description": "...", \\
\quad "paragraph\_context": "..." \\
\} \\
\\
{[Utterance Text]}: "\{UTT\_TEXT\}" \\
{[Emotion Label]}: "\{EMOTION\}" \\
{[Target Language]}: "\{LANG\}"
\end{tcolorbox}
\end{tcolorbox}

\captionof{figure}{Prompt of generating scenario context conditioned on text content and emotion label for scenario-
aware evaluation.}
\label{prompt_for_scene_generation}
\end{center}

%% file: appendix/evaluation_dimension.tex
\section{Task-specific Evaluation Dimensions}
\label{app:task_dimensions}

Table~\ref{tab:task_dimensions} summarizes the evaluation dimensions required by each task in our dataset.
Preference-based tasks output per-dimension scores along with a final A/B decision, while the MOS-style task outputs
seven aspect scores.

% preamble:
% \usepackage{booktabs}
% \usepackage{multirow}

\begin{table*}[tb]
\centering
% \small
\setlength{\tabcolsep}{8pt}
\renewcommand{\arraystretch}{1.1}
\begin{tabular}{c c c}
\toprule
\textbf{Task} & \textbf{Dimension} & \textbf{Range} \\
\midrule
    
\multirow{4}{*}{\parbox{0.25\linewidth}{Task 1:Utterance-Level Speech A/B Preference Judgment}}
& Text Fidelity \& Intelligibility & 0--10 \\
& Speaker Similarity to Prompt Speech & 0--10 \\
& Prosody \& Expressiveness Appropriateness & 0--10 \\
& Naturalness \& Audio Quality & 0--10 \\
\midrule

\multirow{7}{*}{\parbox{0.25\linewidth}{Task 2: Utterance-Level Speech Quality Assessment}}
& Noise & 1--5 \\
& Distortion & 1--5 \\
& Speed (speaking rate) & 1--5 \\
& Continuity (smoothness / discontinuity) & 1--5 \\
& Naturalness & 1--5 \\
& Listening effort & 1--5 \\
& Overall quality & 1--5 \\
\midrule

\multirow{3}{*}{\parbox{0.25\linewidth}{Task 3: Scenario-Aware Style Coherency conditioned on Textual Context}}
& Text Fidelity \& Intelligibility & 0--10 \\
& Scenario Style Match & 0--10 \\
& Naturalness \& Audio Quality & 0--10 \\
\midrule

\multirow{5}{*}{\parbox{0.25\linewidth}{Task 4: Multi-Turn Dialogue Speech Evaluation conditioned on Dialogue History}}
& Intent Matching \& Dialogue Act & 0--10 \\
& Speaker Consistency & 0--10 \\
& Contextual Consistency & 0--10 \\
& Emotion \& Prosody Match & 0--10 \\
& Overall Naturalness & 0--10 \\
\bottomrule
\end{tabular}
\caption{Task-specific evaluation dimensions used in \textsc{UniSRM}.}
\label{tab:task_dimensions}
\end{table*}

\begin{figure*}[t]
    \centering
    % \vspace{-5mm}
    \includegraphics[width=.77\textwidth]{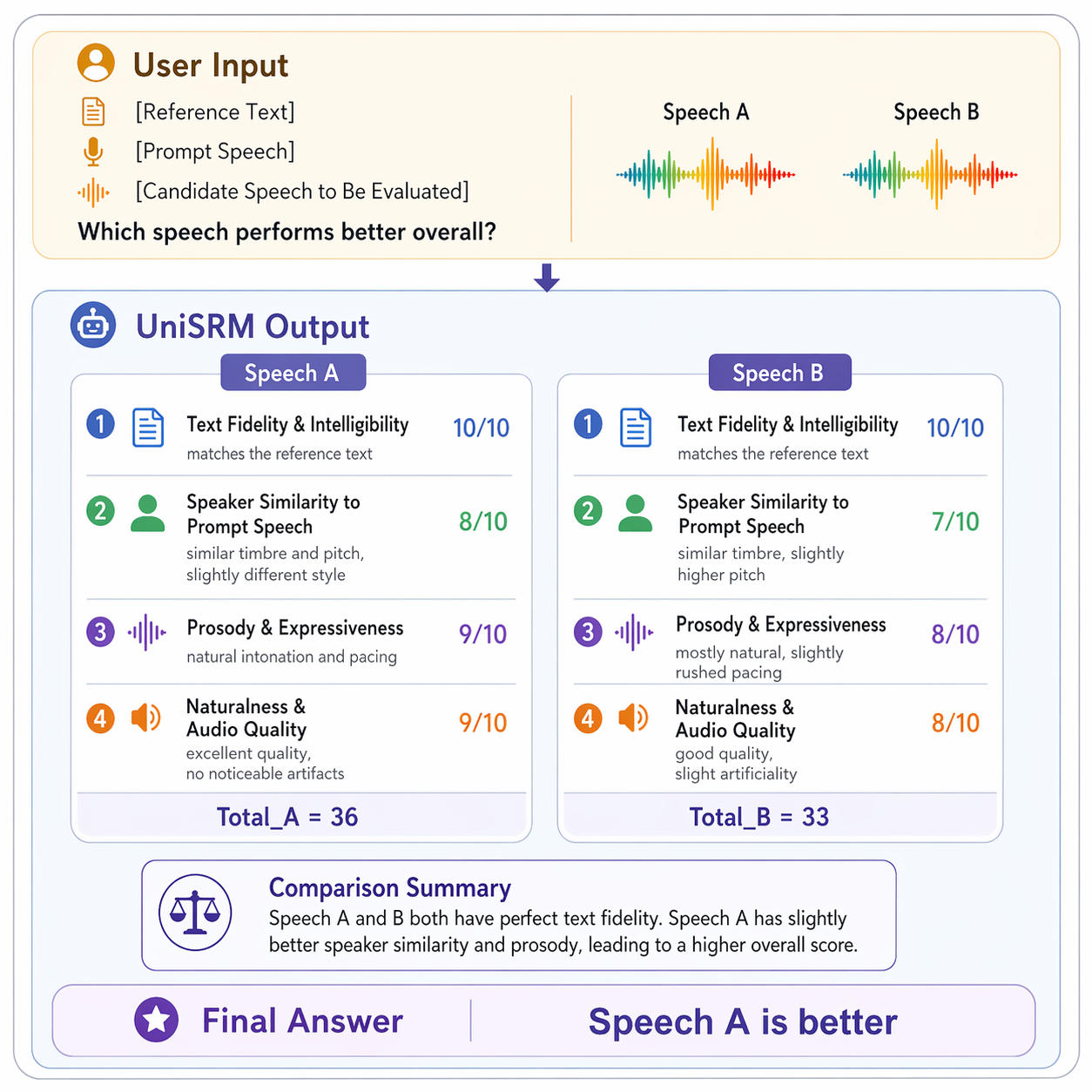}
    \caption{Example output of Task 1 (utterance-level speech A/B preference judgment) in UniSRM.}
    \label{fig:demo_t1}
\end{figure*}

\begin{figure*}[t]
    \centering
    % \vspace{-5mm}
    \includegraphics[width=.8\textwidth]{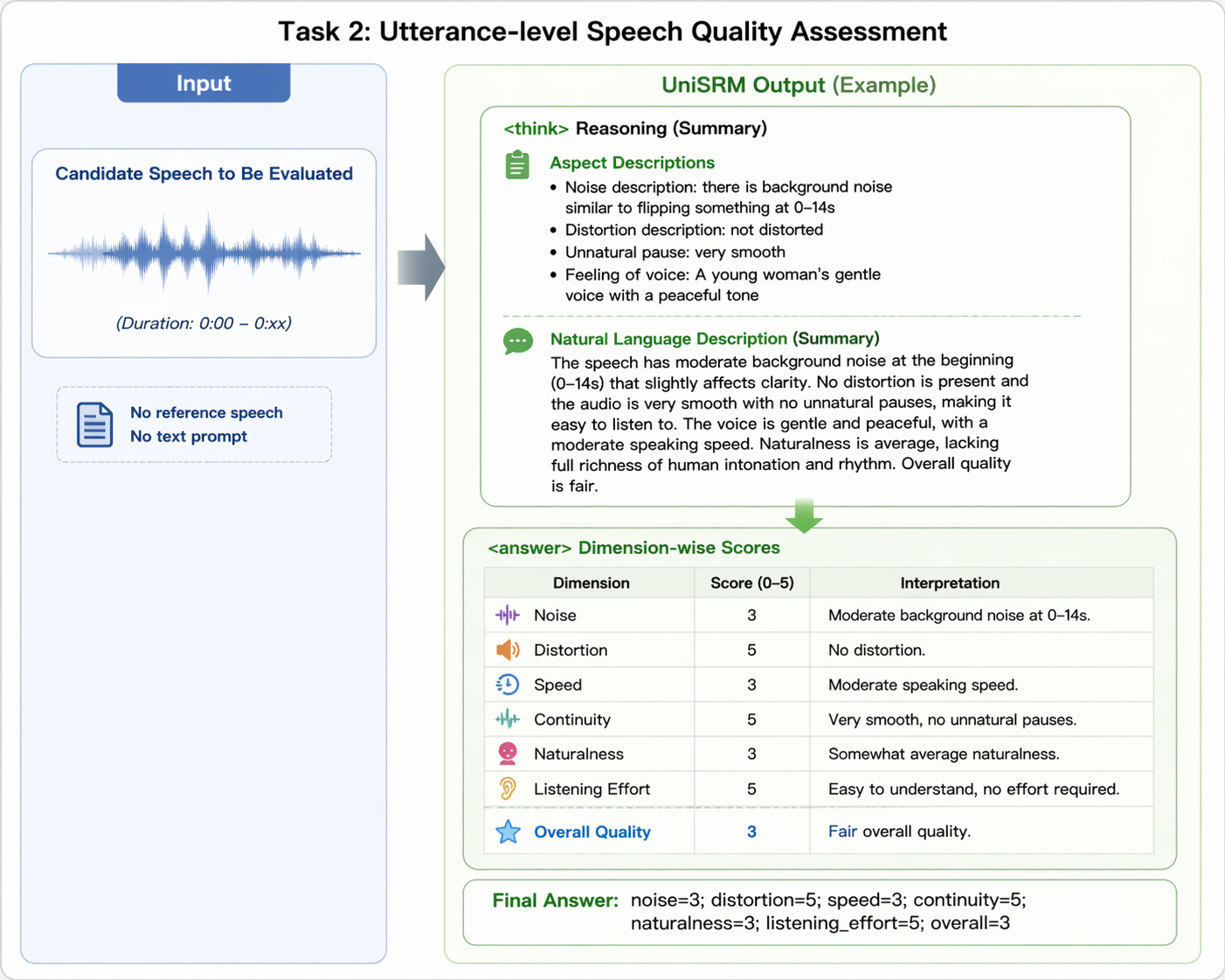}
    \caption{Example output of Task 2 (utterance-level speech quality assessment) in UniSRM.}
    \label{fig:demo_t2}
\end{figure*}

\begin{figure*}[t]
    \centering
    % \vspace{-5mm}
    \includegraphics[width=.77\textwidth]{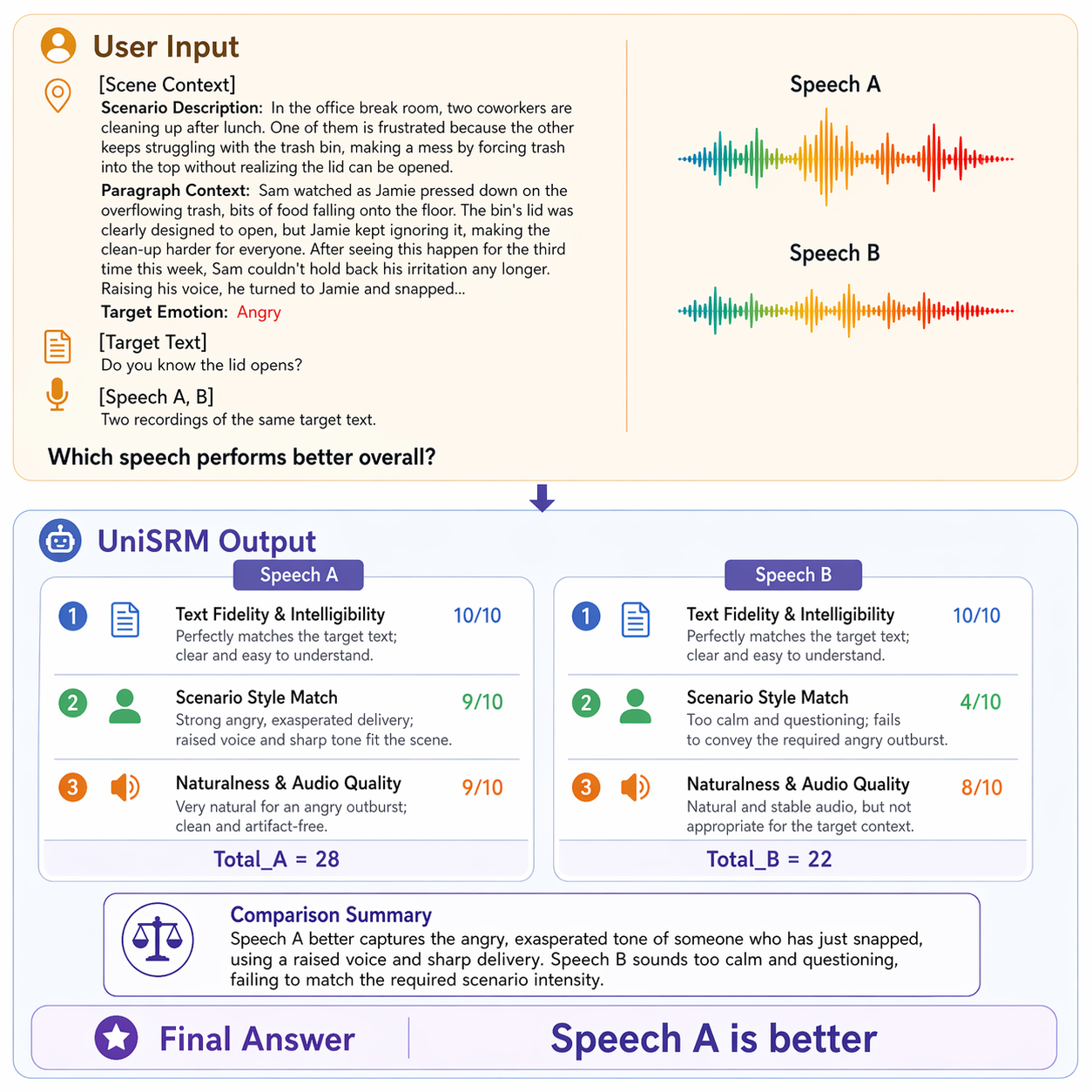}
    \caption{Example output of Task 3 (scenario-aware style
consistency evaluation) in UniSRM.}
    \label{fig:demo_t3}
\end{figure*}

\begin{figure*}[t]
    \centering
    % \vspace{-5mm}
    \includegraphics[width=.77\textwidth]{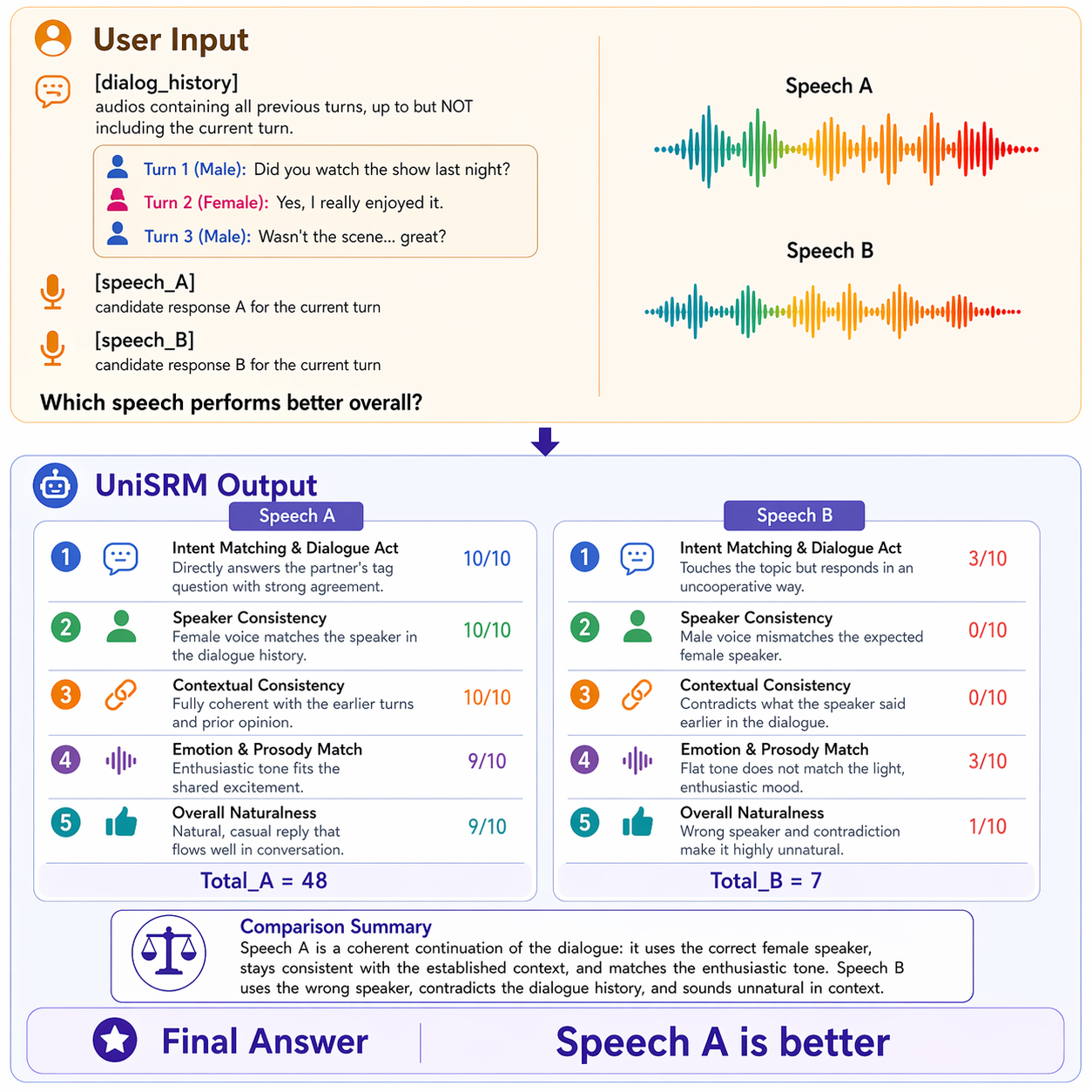}
    \caption{Example output of Task 4 (multi-turn dialogue evaluation) in UniSRM.}
    \label{fig:demo_t4}
\end{figure*}

%% file: appendix/response_demo.tex
\section{Case Study: Example Output}
Figures~\ref{fig:demo_t1}--\ref{fig:demo_t4} present example responses of UniSRM on Task 1 to Task 4, respectively.
The example illustrates how the model conducts a comprehensive multi-dimensional evaluation across different key aspects. 
As shown in the detailed reasoning process, the model provides interpretable, fine-grained assessments for each dimension before aggregating the scores to reach a final comparative judgment.

\input{tables/dataset}

%% file: tables/dataset.tex
\begin{table*}[t]
\centering
% \small
\setlength{\tabcolsep}{7pt}
\renewcommand{\arraystretch}{1.08}
\begin{tabular}{lrrrr}
\toprule
\textbf{Task} & \textbf{SFT} & \textbf{RL} & \textbf{Bench} & \textbf{Total} \\
\midrule
Task 1 (A/B Pref.)        & 11146 & 2787 &  498 & 14431 \\
Task 2 (QualiSpeech MOS)  & 10558 & 2167 & 1852 & 14577 \\
Task 3 (Scenario, EN)     & 5018 & 1815 &  542 & 7375 \\
Task 3 (Scenario, ZH)     &  4869 & 1989 &  506 &  7364 \\
Task 4 (Dialogue)         &  1470 &  916 &  126 &  2512 \\
\midrule
\textbf{Total}            & \textbf{33061} & \textbf{9674} & \textbf{3524} & \textbf{46259} \\
\bottomrule
\end{tabular}
\caption{Data statistics (\#number) of all UniSRM datasets across tasks. Columns report the sizes of the SFT subset, RL subset, and \textsc{UniSRM-Bench}.}
\label{tab:stage_data_stats}
\end{table*}

%% file: appendix/model_config.tex
\section{Model Configuration}
% All training is performed on eight A100 GPUs. 
We use a learning rate of \(1.0\times10^{-5}\) with gradient accumulation of 8 steps for SFT, and a learning rate of \(1.0\times10^{-6}\) with gradient accumulation of 2 steps for RL.
During GRPO, we sample \(G=8\) completions per prompt.
We set the reward weights to \(\lambda_{\text{fmt}}=\lambda_{\text{acc}}=\lambda_{\text{rc}}=1\) when computing the total reward, and use a KL coefficient of \(\beta=0.04\).
For reproducibility, we summarize the main training hyperparameters of both stages in Table~\ref{tab:train_config}. 
We adopt representative open-source TTS systems as speech synthesizers, including CosyVoice2~\cite{du2024cosyvoice}, F5-TTS~\cite{chen2025f5}, ChatTTS\footnote{\url{https://github.com/2noise/ChatTTS}}, and XTTS\footnote{\url{https://github.com/coqui-ai/TTS}}.

\begin{table}[t]
\centering
\begin{tabular}{ccc}
\toprule
 & Stage 1 & Stage 2 \\
\midrule
Batch size / GPU      & 1   & 1 \\
\#GPUs                & 8   & 8 \\
Gradient accumulation & 8   & 2 \\
Effective batch size  & 64  & 16 \\
Learning rate         & $1\times10^{-5}$ & $1\times10^{-6}$ \\
Precision             & bf16 & bf16 \\
\bottomrule
\end{tabular}
\caption{Training configuration of UniSRM.}
\label{tab:train_config}
\end{table}

Objective metrics such as WER, SIM, UTMOS, and DNSMOS are substantially faster to compute, but each captures only a narrow aspect of speech quality and cannot serve as a unified, context-aware evaluator. 
For transparency, we also report the computational cost of UniSRM. 
At inference time, UniSRM runs at 8.98 seconds per iteration with approximately 20\,GB peak GPU memory, which is comparable to SpeechJudge under the same hardware and input settings. 
For training, the SFT stage takes about 4 hours for one epoch on 8 GPUs, corresponding to 30.94 GPU-hours, with roughly 40\,GB peak memory per GPU. 
The GRPO stage takes about 60 hours, corresponding to 480 GPU-hours, with roughly 30\,GB peak memory per GPU.

%% file: appendix/dataset.tex
\section{Dataset statistics}
Building on the data introduction in Section~\ref{sec:data}, Figure~\ref{fig:data_detail} presents the detailed pipeline of UniSRM dataset construction across all tasks.
Task~1 uses a large-scale speech preference dataset constructed from the LibriTTS-R~\cite{koizumi2023librittsr} corpus.
Task~2 is built upon the public QualiSpeech~\cite{wang2025qualispeech} dataset for fine-grained MOS-style speech quality assessment.
Task~3 is constructed from the Emotional Speech Dataset (ESD)~\citep{zhou2022esd} for scenario-aware style consistency under textual context (English and Chinese).
Task~4 is based on DailyTalk~\citep{lee2022dailytalk} for multi-turn dialogue speech evaluation conditioned on dialogue history.
For all tasks, we partition the collected data into SFT, RL, and Bench subsets, ensuring no overlap across splits to avoid leakage.
Finally, the size of datasets used in each training stage is as shown in Table~\ref{tab:stage_data_stats}.

%% file: appendix/eg_exp.tex
\section{Evidence Groundedness (EG) Analysis}

We additionally report EG\_mean to assess the evidence groundedness and reliability of the model’s reasoning, since accuracy-only GRPO ($w/o$ RCR-GRPO) may yield decisions with generic or weak justifications. 
For each sample, we use GPT-4.1 to judge reasoning results of $w/o$ RCR-GRPO and our UniSRM, and then assign an EG score in $\{0,1,2\}$, where $0$ indicates no concrete evidence, $1$ indicates some task-relevant evidence but limited linkage, and $2$ indicates multiple concrete, observable, task-relevant observations clearly supporting the final decision. We then average the scores over all samples to obtain EG\_mean.
Table~\ref{tab:eg_mean} summarizes EG\_mean across tasks, where higher values indicate more evidence-grounded rationales with clearer support for the final choice.
Overall, UniSRM consistently achieves higher EG\_mean, suggesting improved alignment between the reasoning process and the final decision, which can be attributed to the effectiveness of our RCR-GRPO strategy.

\begin{table}[t]
\centering
\small
\begin{tabular}{lccc}
\toprule
\textbf{Model} & \textbf{Task1} & \textbf{Task3-En} & \textbf{Task4} \\
\midrule
$w/o$ RCR-GRPO & 1.28 & 1.66 & 1.95 \\
UniSRM (Ours)      & 1.57 & 1.90 & 1.98 \\
\bottomrule
\end{tabular}
\caption{EG\_mean ($\uparrow$) across tasks.}
\label{tab:eg_mean}
\end{table}

%% file: appendix/human_annotation.tex
\section{Data Quality Control and Human Verification Criteria}
\label{app:data_control}

\subsection{Data Quality Control}
To prevent the reward model from memorizing positional cues, we randomly shuffle the order of $(s_A, s_B)$ for every pair,
so that the first and second positions carry no deterministic meaning.
Since automatic judgments may still contain noise, we further remove self-contradictory samples
by filtering cyclic conflicts, e.g., $A>B,\, B>C,\, C>A$ within the same text group.

To avoid positional bias, we randomly permute the order of $(s_A, s_B)$ for each comparison pair,
so that the first and second positions carry no deterministic meaning.
Moreover, since LLM-based annotations can be noisy, we perform a consistency-based filtering step within each text group:
we discard samples that participate in cyclic contradictions (e.g., $A>B$, $B>C$, yet $C>A$),
which typically indicate unreliable or unstable judgments.
These procedures, together with the human verification criteria described in Appendix~\ref{app:human_criteria},
help improve the overall label reliability for optimization and evaluation.

\subsection{Human Verification Criteria}
\label{app:human_criteria}
To maximize label reliability for optimization and evaluation, we additionally conduct human verification on the GRPO and Test subsets, retaining only samples aligned with human preference.
Before starting the task, each annotator was shown an instruction page describing (i) what data they would listen to, (ii) the purpose of the study (training and evaluating speech reward models), and (iii) how the collected labels would be used (research-only optimization and benchmarking).
Annotators provided informed consent by explicitly agreeing to the task terms prior to annotation.
We paid approximately RMB~1.05 for each pairwise comparison item, i.e., one A/B judgment.
Regarding the underlying speech data, we curate speech samples from publicly available or properly licensed datasets; we follow the corresponding licenses/terms of use, and do not collect additional personal information beyond what is already contained in the source datasets.

Specifically, we perform human verification on the GRPO and Test subsets of utterance-level speech A/B preference judgment task to ensure high label reliability for optimization and evaluation.
Each item presents the same input condition, including target text, speech prompt (reference speaker) and a candidate pair $(s_A, s_B)$,
together with an auto-generated preference label.

% \paragraph{Annotator recruitment and eligibility.}
Each pair is evaluated by three independent annotators.
Annotators must satisfy the following requirements:
\begin{itemize}[leftmargin=1.5em, nosep]
    \item Language proficiency: fluent listening ability in English.
    \item Equipment and environment: headphones are required; annotation must be conducted in a quiet environment.
    \item Hearing and attention: no known hearing impairments.
    \item Qualification: annotators must pass a short qualification test containing gold-reference items before annotation.
\end{itemize}

% \paragraph{Annotation interface and rules.}
For each pair, the interface displays the necessary inputs, including target text, prompt speech, and provides candidate speeches $s_A$ and $s_B$.
Annotators follow the rules below:
\begin{itemize}[leftmargin=1.5em, nosep]
    \item Playback: listen to both samples in full. Replaying is allowed; however, each sample should not be replayed excessively
    (recommended $\leq 3$ times) to avoid fatigue-induced inconsistency.
    \item Order control: the presentation order of $s_A$ and $s_B$ is randomized per annotator.
    \item Decision: provide a binary preference (\textit{A is better} vs.\ \textit{B is better}); ties are not allowed.
    \item Criteria: text fidelity/intelligibility, speaker similarity,
    prosody/expressiveness, and naturalness/audio quality.
    \item Invalid cases: if the pair is not comparable due to corruption, severe truncation, missing content, or other fatal issues,
    mark it as \textit{invalid}.
\end{itemize}

\paragraph{Majority vote and acceptance criteria.}
Let $y^{\text{human}} \in \{A, B, \text{invalid}\}$ denote each annotator's decision.
We retain a pair if and only if it satisfies all of the following:
\begin{itemize}[leftmargin=1.5em, nosep]
    \item Validity: at least two annotators do \emph{not} mark the pair as invalid.
    \item Majority agreement: at least 2 out of 3 annotators agree on the preferred sample.
    \item Consistency with the auto label: the auto-generated preference label by Gemini matches the human majority decision.
\end{itemize}
Pairs failing any criterion are discarded from the GRPO/Test subsets.

\paragraph{Tie and low-consensus handling.}
If the three annotators do not yield a strict majority (e.g., one votes $A$ and one votes $B$ and one marks invalid, or $A/B$ split without majority),
the pair is removed. This conservative filtering avoids introducing ambiguous supervision during GRPO and ensures a clean evaluation set.

\paragraph{Quality control.}
Annotators who consistently fail the gold items or exhibit abnormal behavior (e.g., extremely short annotation time,
always selecting the left item) are excluded, and their annotations are discarded and replaced with re-collected annotations from qualified annotators.

%% file: appendix/multi-turn.tex
\section{Multi-turn dialogue setting}
Figure~\ref{fig:multi_turn_example} shows an example input for the multi-turn dialogue evaluation task (Task~4). 
Both UniSRM-Data and UniSRM-Bench contain multi-turn spoken dialogue samples designed to assess context-dependent judgment ability. 
Each sample includes the full preceding dialogue context and the current response turn to be evaluated. 
For this task, the dialogue history is provided entirely as raw audio, without transcripts, and the full conversation context is used in both training and evaluation.

\begin{center}
\begin{tcolorbox}[
    breakable,
    colback=lavenderback, 
    colframe=lavenderframe,
    arc=1mm,
    boxrule=0.8pt,
    left=5pt, right=5pt, top=5pt, bottom=5pt,
    fonttitle=\bfseries,
    title=Example of a Multi-turn Dialogue Evaluation Instance
]
\small
\setlist[description]{leftmargin=1em, style=nextline}

\textbf{System:} task definition and evaluation rubric.

\smallskip
\textbf{User content (audio sequence):}
\begin{description}[nosep, leftmargin=1.5em]
    \item[{[Turn 1 audio: Speaker A]}] ``Hi, could you help me book a hotel for tomorrow?''
    \item[{[Turn 2 audio: Speaker B]}] ``Sure---what city and budget?''
    \item[{[Turn 3 audio: Speaker A]}] ``Singapore, around 150 SGD.''
\end{description}

\smallskip
\textbf{Target to judge:}
\begin{description}[nosep, leftmargin=1.5em]
    \item[{[Turn 4 audio: Speaker B]}] the response audio to be evaluated.
\end{description}

\end{tcolorbox}
\captionof{figure}{Input example of a multi-turn dialogue evaluation instance.}
\label{fig:multi_turn_example}
\end{center}

%% file: appendix/input-parity.tex
\section{Baseline input parity}

To ensure fair comparisons, we enforce input parity across UniSRM and all judge-style baselines whenever the baseline supports the corresponding modality. The detailed input configuration for each task is summarized in Table~\ref{tab:input_parity}. 
For single-turn tasks, all judge models receive the same audio for evaluation and the same textual task instruction. 
For multi-turn dialogue tasks, all judge models also receive the same raw audio dialogue history, including the full conversation context. 
We do not provide any other side information, such as transcripts or attribute tags, to UniSRM or any baseline. 
For objective metrics that inherently do not support textual prompts or dialogue history (e.g., WER, SIM, UTMOS, DNSMOS), we report results under their native audio-only setting.

\begin{table*}[t]
\centering
\begin{tabular}{cccc}
\toprule
\textbf{Method} & \textbf{Audio to Evaluate} & \textbf{User Prompt} & \textbf{Dialogue History} \\
\midrule
\multicolumn{4}{c}{\textit{Single-turn tasks (T1/T2/T3)}} \\
\midrule
Objective Metrics    & raw audio & \xmark & N/A \\
Proprietary Models   & raw audio & text   & N/A \\
Open-source Models   & raw audio & text   & N/A \\
UniSRM               & raw audio & text   & N/A \\
\midrule
\multicolumn{4}{c}{\textit{Multi-turn dialogue task (T4)}} \\
\midrule
Objective Metrics    & raw audio & \xmark & \xmark \\
Proprietary Models   & raw audio & text   & raw-audio history (full conversation) \\
Open-source Models   & raw audio & text   & raw-audio history (full conversation) \\
UniSRM               & raw audio & text   & raw-audio history (full conversation) \\
\bottomrule
\end{tabular}
\caption{Input parity across UniSRM and baselines. N/A indicates that the corresponding field (e.g., Dialogue History) is not part of the task input. \xmark \ indicates that objective metrics only support audio-only input and do not take textual prompts.}
\label{tab:input_parity}
\end{table*}

\begin{figure*}[t]
    \centering
    % \vspace{-5mm}
    \includegraphics[width=1.0\textwidth]{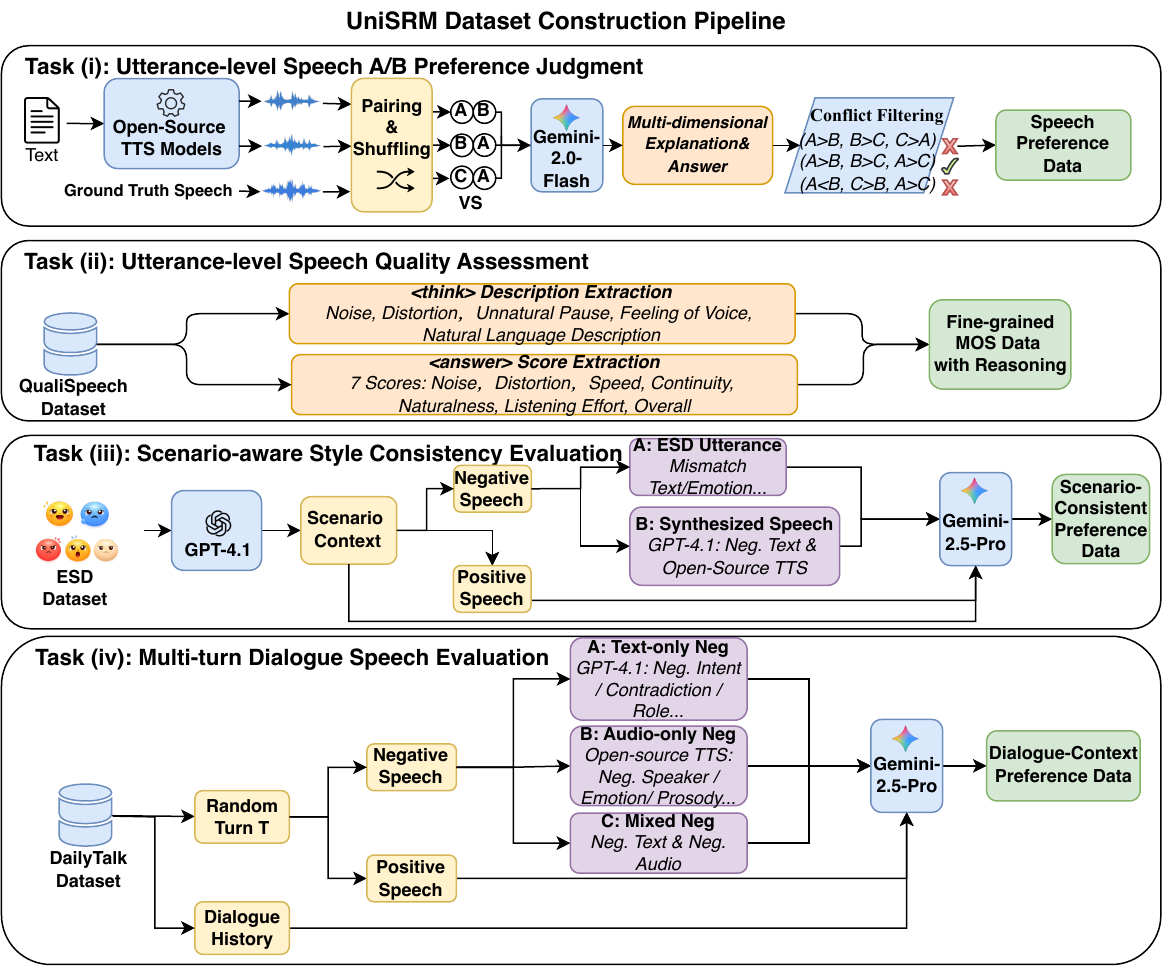}
    \caption{The Detailed Pipeline of UniSRM-Data Construction.} 
    \label{fig:data_detail}
\end{figure*}